\documentclass[12pt,authoryear]{scrartcl}
\usepackage[latin9]{inputenc}
\usepackage[a4paper]{geometry}
\usepackage{amsmath}
\usepackage{graphicx}
\usepackage{wasysym}
\usepackage[numbers]{natbib}
\usepackage[unicode=true,
 bookmarks=false,
 breaklinks=false,pdfborder={0 0 0},pdfborderstyle={},backref=false,colorlinks=false]
 {hyperref}
\hypersetup{
 pdfborderstyle={},pdfborderstyle={}}
\usepackage{breakurl}

\makeatletter

\providecommand{\tabularnewline}{\\}

\usepackage{lineno}
\modulolinenumbers[2]

\usepackage[auth-lg]{authblk}
\author[1]{Tom\'{a}s A. Revilla\footnote{corresponding author, email: tomrevilla@gmail.com}}
\author[1,2]{Vlastimil K\v{r}ivan\footnote{email: vlastimil.krivan@gmail.com}}
\affil[1]{Institute of Entomology, Biology Center, Czech Academy of Sciences, 
                Brani\v{s}ovsk\'{a} 31, 370 05 \v{C}esk\'{e} Bud\v{e}jovice, Czech Republic}
\affil[2]{Department of Mathematics and Biomathematics, Faculty of Science, University of South Bohemia, 
                Brani\v{s}ovsk\'{a} 31, 370 05 \v{C}esk\'{e} Bud\v{e}jovice, Czech Republic}

\makeatother

\begin{document}

\title{Competition, trait\textendash mediated facilitation, and the structure of plant\textendash pollinator
communities}

\date{~}
\maketitle
\begin{abstract}
In plant\textendash pollinator communities many pollinators are potential generalists
and their preferences for certain plants can change quickly in response to changes
in plant and pollinator densities. These changes in preferences affect coexistence
within pollinator guilds as well as within plant guilds. Using a mathematical model,
we study how adaptations of pollinator preferences influence population dynamics
of a two-plant\textendash two-pollinator community interaction module. Adaptation
leads to coexistence between generalist and specialist pollinators, and produces
complex plant population dynamics, involving alternative stable states and discrete
transitions in the plant community. Pollinator adaptation also leads to plant\textendash plant
apparent facilitation that is mediated by changes in pollinator preferences. We show
that adaptive pollinator behavior reduces niche overlap and leads to coexistence
by specialization on different plants. Thus, this article documents how adaptive
pollinator preferences for plants change the structure and coexistence of plant\textendash pollinator
communities.

\end{abstract}
\begin{quote}
\emph{The pedigree of honey}

\emph{Does not concern the bee;}

\emph{A clover, any time, to him}

\emph{Is aristocracy.}

Poems (1890) \textendash{} Emily Dickinson
\end{quote}

\section{Introduction}

Many mutualistic interactions feature direct resource-for-resource (e.g., plant\textendash mycorrhizae,
lichens), or resource-for-service (e.g., pollination, seed dispersal) exchanges between
species, but this fact was not explicitly considered by the first models of mutualism
based on the Lotka\textendash Volterra equations \citep{gause_witt-amnat35,vandermeer_boucher-jtb78}.
As a result, positive feedbacks between mutualists predicted infinite population
growth. Later models considered negative density dependence at high population densities
\citep{boucher1988,hernandez-rspb98,gerla_mooij-tpb14} that stabilizes population
dynamics. Increased awareness about the consumer\textendash resource aspects of mutualisms
\citep{holland_deangelis-ecology10} provides some mechanistic underpinnings for
density dependence (e.g., mutualistic benefits saturate, just like plant growth saturates
with nutrients or predator feeding saturates with prey). More recently, differentiation
between non-living mutualistic resources (e.g., mineral nutrients, nectar, fruits)
and their living providers (e.g., fungi, plant) led to several mechanistic models
\citep{benadi_etal-amnat12,valdovinos_etal-oikos13,revilla-jtb15}. These are very
relevant for studies of plant\textendash animal mutualisms, like pollination and
seed dispersal, for two reasons. First, competition between animals for nectar or
fruits can be treated using concepts from consumer\textendash resource theory \citep{grover1997}.
Second, competition between plants for pollination or seed dispersal can result from
plants influencing the preferences of animals, according to optimal foraging theory
\citep{pyke-ppees16}.

In an earlier work \citep{revilla_krivan-plosone16} we analyzed coexistence conditions
for two plants competing for a single pollinator. If the pollinator is a generalist,
plants can facilitate each other by making the pollinator more abundant. Facilitation
is an example of an indirect density-mediated interaction \citep[sensu][]{bolker_etal-ecology03}
between the two plants. However, if pollinators have adaptive preferences, a positive
feedback between plant abundance and pollinator preferences predicts exclusion of
the rare plant, which gets less pollination as pollinators specialize on the common
plant. In other words, when pollinator preferences respond to plant densities, plants
will experience competition for pollination services (in addition to competition
for other factors such as nutrients, light or space) because an increase in pollination
of one plant exerts a negative effect on the other plants that gets less pollination.
In \citet{revilla_krivan-plosone16} we found that plant coexistence depends on the
balance between plant facilitation via increasing abundance of the common pollinator,
and competition for pollinator preferences, which adapt in response to the relative
abundance of plant resources. Pollinator preferences were described by the ideal
free distribution \citep[IFD;][]{fretwell_lucas-actabiot69} that predicts pollinator
distribution between the two plants in such a way that neither of the two plants
provides pollinators with a higher payoff. For a single pollinator, the IFD is also
an evolutionarily stable strategy \citep[ESS,][]{krivan_etal-tpb08}, i.e., once
adopted by all individuals no mutant with a different strategy can invade the resident
population \citep{smith_price-nature73}.

In many real life settings however, plants compete for pollination services provided
by several pollinator species, which in turn compete for plant resources. Pollinator
preferences for plants respond not only to plant abundances, but also to inter- and
intra-specific competition between pollinators. Simulations of large plant\textendash pollinator
communities indicate that plant coexistence is promoted when generalist pollinators
specialize to reduce competition for resources, i.e., to decrease niche overlap \citep{valdovinos_etal-oikos13,valdovinos_etal-ecolett16}.
This is the classic competitive exclusion principle which states that $n$ competing
species (i.e., pollinators) cannot coexist at a population equilibrium if they are
limited by less than $n$ limiting factors (i.e., plants) \citep{levin-amnat70}.

In this article we study a mutualistic\textendash competitive interaction module
consisting of two plants and two pollinators where pollinators behave as adaptive
foragers that maximize their fitness depending on plant resource quality and abundance.
This means that depending on plant and pollinator densities, pollinators switch between
generalism and specialism. These behavioral changes also change the topology of the
interaction network. Thus, we focus on two questions: Under what conditions the two
plants and two pollinators can coexist at an equilibrium, and what are the corresponding
community network configurations.

To gain insight, we study separately plant population dynamics at fixed pollinator
densities, and pollinator population dynamics at fixed plant densities, respectively.
In both cases we compare population dynamics for inflexible pollinators with those
for adaptive pollinators. Under fixed pollinator preferences (section \ref{sec:FixPref}),
stable coexistence of plants, or pollinators, is possible at a unique equilibrium.
It is also possible that at this population equilibrium both pollinators are generalists.
Both these predictions change when pollinator preferences for plants are adaptive
(section \ref{sec:AdPref}). First, when pollinator densities are fixed, plants can
coexist at alternative stable states characterized by different interaction topologies
given by pollinator strategy. However, there is no plant stable coexistence when
both pollinators are generalists. Second, when plant densities are fixed, pollinators
can coexist at an equilibrium only if they specialize on different plants (section
\ref{subsec:polli_comp_adpref}). We show how these conclusions can explain some
recent experimental and simulated results, as well as predict the effects of pollinator
adaptation in real communities.

\section{Population dynamics when pollinator preferences for plants are fixed\label{sec:FixPref}}

Consider two plant populations P1 and P2 interacting with two pollinator populations
A1 and A2. Mutualism is mediated by resources R1 and R2 produced by plants P1 and
P2, respectively. We assume that pollination is concomitant with pollinator resource
consumption. Since resources like nectar or pollen have much faster turnover dynamics
(hours, days) than plants and pollinators (weeks, months), we assume they attain
a quasi-steady-state at current plant and animal densities \citep{revilla-jtb15}.
As a result, population dynamics follow the \citet{revilla_krivan-plosone16} model
for a single pollinator, extended for two pollinators

\begin{subequations}\label{eq:ode_system}

\begin{align}
\frac{dP_{1}}{dt} & =\bigg(\frac{a_{1}(r_{11}u_{1}b_{11}A_{1}+r_{12}v_{1}b_{12}A_{2})}{w_{1}+u_{1}b_{11}A_{1}+v_{1}b_{12}A_{2}}\left(1-\frac{P_{1}+c_{2}P_{2}}{K_{1}}\right)-m_{1}\bigg)P_{1}\label{eq:ode_p1}\\
\frac{dP_{2}}{dt} & =\bigg(\frac{a_{2}(r_{21}u_{2}b_{21}A_{1}+r_{22}v_{2}b_{22}A_{2})}{w_{2}+u_{2}b_{21}A_{1}+v_{2}b_{22}A_{2}}\left(1-\frac{P_{2}+c_{1}P_{1}}{K_{2}}\right)-m_{2}\bigg)P_{2}\label{eq:ode_p2}\\
\frac{dA_{1}}{dt} & =\bigg(\frac{a_{1}e_{11}u_{1}b_{11}P_{1}}{w_{1}+u_{1}b_{11}A_{1}+v_{1}b_{12}A_{2}}+\frac{a_{2}e_{21}u_{2}b_{21}P_{2}}{w_{2}+u_{2}b_{21}A_{1}+v_{2}b_{22}A_{2}}-d_{1}\bigg)A_{1}\label{eq:ode_a1}\\
\frac{dA_{2}}{dt} & =\bigg(\frac{a_{1}e_{12}v_{1}b_{12}P_{1}}{w_{1}+u_{1}b_{11}A_{1}+v_{1}b_{12}A_{2}}+\frac{a_{2}e_{22}v_{2}b_{22}P_{2}}{w_{2}+u_{2}b_{21}A_{1}+v_{2}b_{22}A_{2}}-d_{2}\bigg)A_{2},\label{eq:ode_a2}
\end{align}

\end{subequations}

\noindent where $P_{i}$ ($i=1,2$) is plant Pi population density, and $A_{j}$
($j=1,2$) is pollinator Aj population density. Here $a_{i}$ is a plant resource
production rate, $w_{i}$ is its spontaneous decay rate, and $b_{ij}$ is a pollinator
specific consumption rate. In the plant equations (\ref{eq:ode_p1},\ref{eq:ode_p2}),
pollinator consumption rates translate into seed production rates with efficiency
$r_{ij}$. Plant growth is reduced by intra-specific competition, with carrying capacity
$K_{i}$, and by inter-specific competition, where $c_{i}$ is the relative effect
of plant $i$ on the other plant. In the absence of pollinators, plants die with
per-capita rates $m_{i}$, so plants are obligate mutualists. In the pollinator equations
(\ref{eq:ode_a1},\ref{eq:ode_a2}), consumption translates into growth with efficiency
ratios $e_{ij}$. Without plants, pollinators die with per-capita rates $d_{j}$,
so pollinators are obligate mutualists too.

Pollinator A1 (A2) preferences are $u_{1}$ $(v_{1})$ for plant P1 and $u_{2}=1-u_{1}$
$(v_{2}=1-v_{1})$ for plant P2. Preferences can be interpreted as fractions of foraging
time that individual pollinators spend on plant P1 or P2, or the proportion of a
pollinator population which is visiting P1 or P2 at a given time. Preferences allows
us to categorize pollinators as generalists or specialists. For example, if $(u_{1},u_{2})=(3/4,1/4)$
and $(v_{1},v_{2})=(0,1)$, then A1 is a generalist (biased towards P1) and A2 is
a P2 specialist. In this section we assume that pollinator preferences for plants
are fixed and we derive conditions for plant stable coexistence that are compared
in section \ref{sec:AdPref} with the case where pollinator preferences are adaptive.
Unfortunately, the many variables and parameters of model (\ref{eq:ode_system})
do not allow us to analyze it at this generality. In order to gain insights, we assume
that either plants or pollinators are kept at fixed densities and employing isocline
analysis \citep{case2000} we characterize coexistence between plants (\ref{eq:ode_p1},\ref{eq:ode_p2}),
or between pollinators (\ref{eq:ode_a1},\ref{eq:ode_a2}).

\subsection{Plant coexistence}

\label{plant_coexistence}

First, we consider plant-only dynamics. Let us consider a community consisting of
a single plant Pi ($i=1,2$) and two pollinators. At fixed pollinator densities $A_{1}$
and $A_{2}$, the necessary condition for plant Pi to survive is that its pollinator-dependent
per-capita birth rate is higher than its mortality rate, i.e., 
\begin{equation}
r_{i}=\frac{a_{i}(r_{i1}u_{i}b_{i1}A_{1}+r_{i2}v_{i}b_{i2}A_{2})}{w_{i}+u_{i}b_{i1}A_{1}+v_{i}b_{i2}A_{2}}>m_{i},\label{eq:r_plv}
\end{equation}

\noindent in which case the plant will attain its pollinator-dependent carrying capacity
\begin{equation}
H_{i}=K_{i}\bigg(1-\frac{m_{i}}{r_{i}}\bigg).\label{eq:h_plv}
\end{equation}

Inequality (\ref{eq:r_plv}) shows that if both pollinators have low preferences
for plant Pi (i.e., both $u_{i}$ and $v_{i}$ are small), the plant cannot achieve
a positive growth rate and cannot invade when rare. To invade, a plant must be attractive
enough for at least one of the two pollinators.

\begin{figure}
\begin{centering}
\includegraphics[width=1\textwidth]{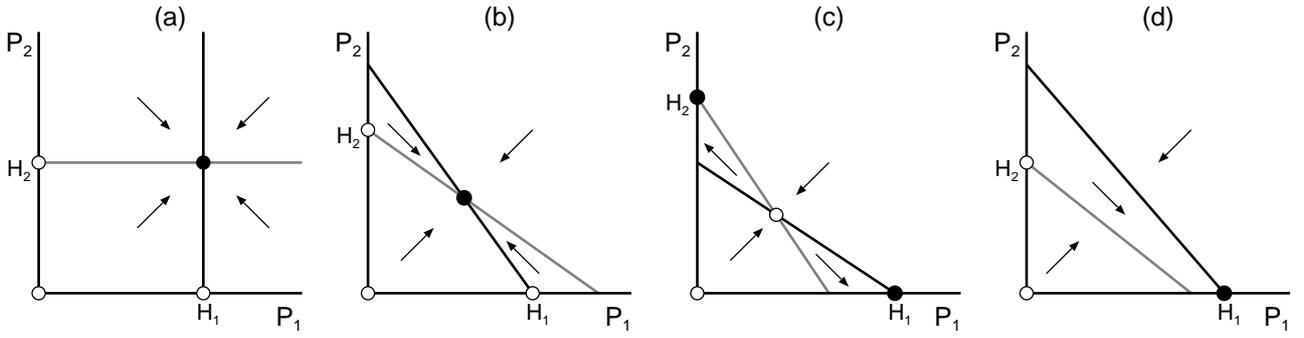} 
\par\end{centering}
\caption{\label{fig:nullclines}Qualitative configurations of plant isoclines (P1 in black
and P2 in gray) when pollinator preferences for plants and densities are fixed. Filled
(open) circles represent stable (unstable) equilibria. Circles on the axes correspond
to pollinator-dependent carrying capacities $H_{i}$ given by (\ref{eq:h_plv}).}
\end{figure}

Provided that (\ref{eq:r_plv}) holds for both plants, the plant sub-system (\ref{eq:ode_p1},\ref{eq:ode_p2})
is the Lotka-Volterra competition model. Plant coexistence depends on inter-specific
competition coefficients $(c_{1},c_{2})$, and the carrying capacities given by (\ref{eq:h_plv}).
Figure \ref{fig:nullclines} shows all generic qualitative plant isocline configurations
and their outcomes for plant coexistence. Panel (a) shows the non-competitive case
$(c_{1}=c_{2}=0)$ where both plants attain their pollinator-dependent carrying capacities
$H_{i}$. Under direct competition $(c_{1},c_{2}>0)$ plant equilibrium densities
at coexistence are lower than $H_{i}$ (panels b, c). If

\begin{equation}
c_{1}<\frac{H_{2}}{H_{1}}\;\;\;\text{and}\;\;\;c_{2}<\frac{H_{1}}{H_{2}},\label{eq:plant_stability}
\end{equation}

\noindent isoclines intersect in the positive quadrant at the globally stable equilibrium
(panel b)

\[
(P_{1},P_{2})=\bigg(\frac{H_{1}-c_{2}H_{2}}{1-c_{1}c_{2}},\frac{H_{2}-c_{1}H_{1}}{1-c_{1}c_{2}}\bigg).
\]

If opposite inequalities hold in (\ref{eq:plant_stability}), the coexistence equilibrium
is unstable (panel c), with one plant outcompeting the other plant depending on the
initial conditions. If the isoclines do not intersect in the first quadrant the species
with the highest (i.e., the one which is above the other) isocline always wins (i.e.,
plant P1 in panel d). The height of a plant's isocline depends on its carrying capacity
$H_{i}$. Given that $H_{i}$ increases with $u_{i}$ and $v_{i}$ (since $r_{i}$
in (\ref{eq:r_plv}) increases with $u_{i}$ and $v_{i}$), the more preferred a
plant is, the more numerous will it be under conditions of stable coexistence, or
more likely it will exclude the other plant.

\subsection{Pollinator coexistence}

Second, we consider pollinator-only dynamics. For fixed plant densities $P_{i}$
($i=1,2$), the pollinator sub-system (\ref{eq:ode_a1},\ref{eq:ode_a2}) is the
resource competition model of \citet{schoener-tpb78}. Appendix \ref{sec:appA1A2coex}
shows that there are three qualitatively different pollinator equilibria. The equilibrium
where both pollinators are extinct $(A_{1},A_{2})=(0,0)$ is unstable if one or both
pollinators is viable. Viability conditions for pollinator A1 and A2 are, respectively,

\begin{subequations}

\begin{align}
a_{1}P_{1}e_{11}u_{1}b_{11}w_{2}+a_{2}P_{2}e_{21}u_{2}b_{21}w_{1} & >d_{1}w_{1}w_{2}\label{viabA1}\\
a_{1}P_{1}e_{12}v_{1}b_{12}w_{2}+a_{2}P_{2}e_{22}v_{2}b_{22}w_{1} & >d_{2}w_{1}w_{2}.\label{viabA2}
\end{align}

\end{subequations}

If neither of the above inequalities holds, both pollinators go extinct. If only
one 
inequality holds then the corresponding pollinator is viable, and for each viable
pollinator there is a corresponding single species equilibrium $(A_{1},0)$ or $(0,A_{2}).$
As we see, pollinator viability implies minimum resource requirements \citep{grover1997}.

Appendix \ref{sec:appA1A2coex} shows that there can be at most one pollinator coexistence
equilibrium $(\hat{A}_{1},\hat{A}_{2})$. Such an equilibrium is locally asymptotically
stable (Appendix \ref{sec:appA1A2coex}) if

\begin{equation}
(u_{1}b_{11}v_{2}b_{22}-v_{1}b_{12}u_{2}b_{21})(e_{11}u_{1}b_{11}e_{22}v_{2}b_{22}-e_{12}v_{1}b_{12}e_{21}u_{2}b_{21})>0.\label{eq:pollinator_stability}
\end{equation}

The interpretation of condition (\ref{eq:pollinator_stability}) is similar to that
given by \citet{leon_tumpson-jtb75} for two consumers competing for two substitutable
resources: \emph{``... the competitors coexist if at equilibrium each of them removes
at a higher rate that resource which contributes more to its own rate of growth.''}
To see why this is so, let us assume that plant P1 is better for the growth of A1
($e_{11}>e_{21}$) and P2 is better for the growth of A2 ($e_{22}>e_{12}$). Then,
if pollinator A1 interacts comparatively more strongly with plant P1 than with P2
($u_{1}b_{11}>u_{2}b_{21}$), and pollinator A2 interacts comparatively more strongly
with plant P2 than with P1 ($v_{2}b_{22}>v_{1}b_{12}$), inequality (\ref{eq:pollinator_stability})
holds.

\begin{figure}
\begin{centering}
\includegraphics[width=1\textwidth]{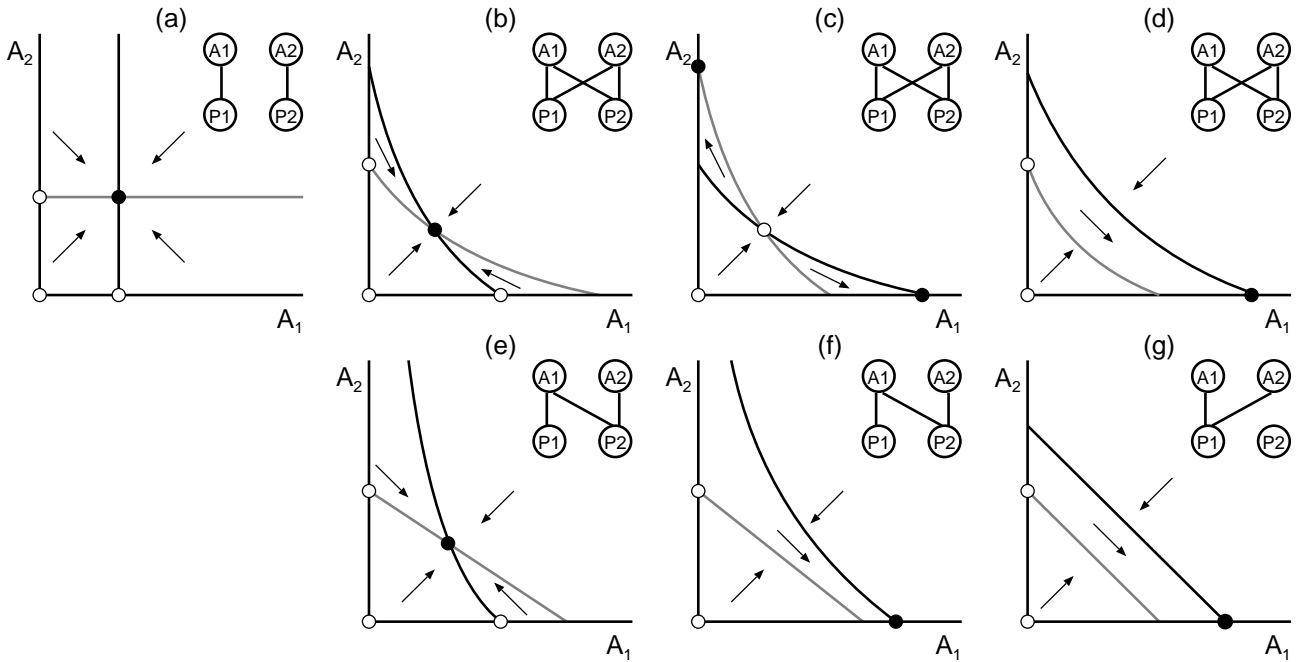} 
\par\end{centering}
\caption{\label{fig:nullclines_pollinators}Pollinator isocline configurations (A1 in black
and A2 in gray) and qualitative dynamics (arrows), at fixed pollinator preferences.
Filled (open) circles represent stable (unstable) equilibria. Isocline shapes depend
on interaction topology (inset graphs).}
\end{figure}

Provided both pollinators are viable (\ref{viabA1} and \ref{viabA2} hold), Figure
\ref{fig:nullclines_pollinators} shows all generic pollinator isocline configurations
corresponding to different interaction topologies (except symmetries). The top row
of this figure is analogous to Figure \ref{fig:nullclines} for plants. Panel (a)
shows the case where pollinators specialize on different plants ($u_{1}=1,v_{1}=0$).
The A1 isocline is vertical, the A2 isocline is horizontal, and their intersection
corresponds to stable pollinator coexistence since pollinators do not compete. Panels
(b,c,d) display isoclines for two generalist pollinators (i.e., $0<u_{1}<1$, $0<v_{1}<1$),
i.e., both pollinators share both plants. Notice that the isoclines of generalist
pollinators are curved and intersect both axes. In (b) an isocline intersection exists
and the equilibrium between generalists is globally stable because (\ref{eq:pollinator_stability})
holds. In (c) an isocline intersection exists but the corresponding equilibrium between
generalists is unstable because (\ref{eq:pollinator_stability}) does not hold and
either A1 or A2 wins the competition depending on the initial conditions. In panel
(d) the isoclines do not intersect and the pollinator with the highest isocline always
wins. In other words condition (\ref{eq:pollinator_stability}) is irrelevant for
coexistence in this case. This outcome happens if e.g., A1 has a much lower mortality
and/or higher conversion efficiencies than A2. This case is like the case of competitive
dominance between plants (Figure \ref{fig:nullclines}d), except that for the plants
the isoclines are linear.

Panels (e,f) display isoclines when pollinator A1 is a generalist and A2 is a P2
specialist (i.e., $0<u_{1}<1$, $v_{1}=0$). Like in panels (b,c,d) the isocline
of the generalist is curved, but the specialist isocline is linear. Under these condition,
condition (\ref{eq:pollinator_stability}) is trivially satisfied (because $v_{1}=0$).
Thus, if both isoclines intersect, the corresponding coexistence equilibrium is always
globally stable like in panel (e), and if they do not intersect the species with
the highest isocline always wins (e.g., A1 in panel (f)). In other words, competition
between a generalist and a specialist pollinator does not admit the bi-stable case
(i.e., panel c).

Finally, in panel (g) both pollinators specialize on plant P1, (e.g., $u_{1}=v_{1}=1$).
In this case both pollinators have parallel linearly decreasing isoclines, and the
pollinator with the higher isocline (i.e., A1 in this case) excludes the other pollinator.
This case is like the case of competitive dominance between plants (Figure \ref{fig:nullclines}d),
except that for the plants the isoclines are not required to be parallel.

\section{Population dynamics when pollinator preferences for plants are adaptive\label{sec:AdPref}}

In this section we assume that pollinator preferences adaptively change as plant
and pollinator densities change. First (section \ref{subsec:opti_pref}), we use
a game theoretic approach \citep{krivan_etal-tpb08} to derive optimal pollinator
preferences at given plant and pollinator densities. Second (section \ref{subsec:plant_comp_adpref}),
we analyze competition between plants at fixed pollinator densities. Third (section
\ref{subsec:polli_comp_adpref}), we analyze competition between pollinators at fixed
plant densities.

\subsection{Optimal pollinator preferences\label{subsec:opti_pref}}

Let us consider a mutant pollinator A1 with preference $\tilde{u}_{1}\in[0,1]$ for
the first plant and a mutant pollinator A2 with preference $\tilde{v}_{1}\in[0,1]$
in a resident population of pollinators with average preferences $u_{1}$ and $v_{1}$,
respectively. The payoff a pollinator obtains when pollinating plant $i$ ($i=1,2$)
is given by the per-capita pollinator birth rate. For example, from (\ref{eq:ode_a1})
the payoff of a pollinator A1 when pollinating plant P2 is $\frac{a_{2}e_{21}b_{21}P_{2}}{w_{2}+u_{2}b_{21}A_{1}+v_{2}b_{22}A_{2}}$.
As the resident pollinator distribution between the two plants is the same as are
their preferences we see that payoffs depend on the distribution of pollinators between
the two plants. Fitnesses of A1 and A2 mutants are defined as their mean payoffs

\begin{subequations}\label{eq:payoff_mutant}

\begin{align}
F_{1}(\tilde{u}_{1};u_{1},v_{1}) & =\frac{a_{1}e_{11}b_{11}P_{1}}{w_{1}+u_{1}b_{11}A_{1}+v_{1}b_{12}A_{2}}\tilde{u}_{1}+\frac{a_{2}e_{21}b_{21}P_{2}}{w_{2}+u_{2}b_{21}A_{1}+v_{2}b_{22}A_{2}}\tilde{u}_{2},\label{eq:fitnessA1mutant}\\[0.5cm]
F_{2}(\tilde{v}_{1};u_{1},v_{1}) & =\frac{a_{1}e_{12}b_{12}P_{1}}{w_{1}+u_{1}b_{11}A_{1}+v_{1}b_{12}A_{2}}\tilde{v}_{1}+\frac{a_{2}e_{22}b_{22}P_{2}}{w_{2}+u_{2}b_{21}A_{1}+v_{2}b_{22}A_{2}}\tilde{v}_{2}.\label{eq:fitnessA2mutant}
\end{align}

\end{subequations}

Throughout the rest of this article we assume that pollinator A1 grows comparatively
faster on plant P1 than on P2, and that pollinator A2 grows comparatively faster
on plant P2 than on P1, i.e.,

\begin{equation}
(e_{11}b_{11})(e_{22}b_{22})>(e_{21}b_{21})(e_{12}b_{12}).\label{eq:pollinator_ess_condition}
\end{equation}

We want to find pollinator preferences for plants that are evolutionarily stable
\citep{hofbauersigmund1998}. Interestingly, Appendix \ref{sec:appESS} shows that
there is no evolutionarily stable preference/strategy where both pollinator species
behave as generalists (i.e., preference $(u_{1},v_{1})$ where $0<u_{1}<1$ and $0<v_{1}<1$).
In other words, the interaction topology in Figure \ref{fig:nullclines_pollinators}b,c,d
does not exist when pollinators preferences are adaptive. In fact either both species
are specialists, or one species is a generalist and the other specializes on the
plant that makes it grow faster. Table \ref{tab:regions_plaspace} lists all possible
ESSs as a function of plant and pollinator population densities. Transitions between
ESSs in plant phase space occur along four lines $P_{2}=Q_{i}P_{1}$ $(i=a,b,c,d)$,
called isolegs \citep{rosenzweig-ecology81,pimm_rosenzweig-oikos81,krivan_sirot-amnat02},
where

\begin{subequations}\label{eq:isolegs}

\begin{align}
Q_{a}(A_{1},A_{2}) & =\frac{a_{1}b_{11}e_{11}(w_{2}+b_{21}A_{1}+b_{22}A_{2})}{a_{2}b_{21}e_{21}w_{1}},\label{eq:isoleg_a}\\[0.2cm]
Q_{b}(A_{1},A_{2}) & =\frac{a_{1}b_{11}e_{11}(w_{2}+b_{22}A_{2})}{a_{2}b_{21}e_{21}(w_{1}+b_{11}A_{1})},\label{eq:isoleg_b}\\[0.2cm]
Q_{c}(A_{1},A_{2}) & =\frac{a_{1}b_{12}e_{12}(w_{2}+b_{22}A_{2})}{a_{2}b_{22}e_{22}(w_{1}+b_{11}A_{1})},\label{eq:isoleg_c}\\[0.2cm]
Q_{d}(A_{1},A_{2}) & =\frac{a_{1}b_{12}e_{12}w_{2}}{a_{2}b_{22}e_{22}(w_{1}+b_{11}A_{1}+b_{12}A_{2})}.\label{eq:isoleg_d}
\end{align}

\end{subequations}

At fixed pollinator densities isolegs delineate five regions (denoted as I-V in Table
\ref{tab:regions_plaspace}) in the first quadrant of the $P_{1}P_{2}$ plane where
pollinators behave as specialists or generalists. Appendix \ref{sec:appESS} shows
that when pollinator A1 is a generalist and A2 specializes on P2 (region II in Table
\ref{tab:regions_plaspace}), the ESS of A1 is

\begin{subequations}\label{eq:ess}

\begin{equation}
u_{1}^{*}=\frac{e_{11}b_{11}a_{1}P_{1}(w_{2}+b_{21}A_{1}+b_{22}A_{2})-e_{21}b_{21}a_{2}P_{2}w_{1}}{b_{11}b_{21}(e_{11}a_{1}P_{1}+e_{21}a_{2}P_{2})A_{1}},\label{eq:u_star}
\end{equation}

\noindent and when A2 is a generalist and A1 specializes on P1 (region IV in Table
\ref{tab:regions_plaspace}), the ESS of A2 is

\begin{equation}
v_{1}^{*}=\frac{e_{12}b_{12}a_{1}P_{1}(w_{2}+b_{22}A_{2})-e_{22}b_{22}a_{2}P_{2}(w_{1}+b_{11}A_{1})}{b_{12}b_{22}(e_{12}a_{1}P_{1}+e_{22}a_{2}P_{2})A_{2}}.\label{eq:v_star}
\end{equation}

\end{subequations}

\begin{table}
\begin{centering}
\begin{tabular}{cccl}
\hline 
Region  & Conditions  & ESS $(u_{1},v_{1})$  & Description\tabularnewline
\hline 
I  & $Q_{a}(A_{1},A_{2})P_{1}<P_{2}$  & $(0,0)$  & A1 \& A2 specialize on P2\tabularnewline
II  & $Q_{b}(A_{1},A_{2})P_{1}<P_{2}<Q_{a}(A_{1},A_{2})P_{1}$  & $(u_{1}^{*},0)$  & A1 generalist, A2 specializes on P2\tabularnewline
III  & $Q_{c}(A_{1},A_{2})P_{1}<P_{2}<Q_{b}(A_{1},A_{2})P_{1}$  & $(1,0)$  & A1 specializes on P1, A2 specializes on P2\tabularnewline
IV  & $Q_{d}(A_{1},A_{2})P_{1}<P_{2}<Q_{c}(A_{1},A_{2})P_{1}$  & $(1,v_{1}^{*})$  & A1 specializes on P1, A2 generalist\tabularnewline
V  & $P_{2}<Q_{d}(A_{1},A_{2})P_{1}$  & $(1,1)$  & A1 \& A2 specialize on P1\tabularnewline
\hline 
\end{tabular}
\par\end{centering}
\caption{\label{tab:regions_plaspace}Dependence of evolutionarily stable pollinator preferences
on plant $(P_{1},P_{2})$ and pollinator densities $(A_{1},A_{2})$. Thresholds $Q_{i}$
$(i=a,b,c,d)$ are given by (\ref{eq:isolegs}) and $u_{1}^{*}$ and $v_{1}^{*}$
by (\ref{eq:ess}).}
\end{table}

In the next section we use isolegs and isoclines to study plant\textendash plant
competition.

\subsection{Plants compete for pollinator preferences\label{subsec:plant_comp_adpref}}

Here we use isocline analysis to study the dynamics of the plant sub-system at fixed
pollinator densities $A_{1}$ and $A_{2}$, when pollinators are adaptive. Unlike
in the case with fixed preferences, pollinator isolegs partition the $P_{1}P_{2}$
plane into five regions listed in Table \ref{tab:regions_plaspace}. Isolegs $P_{2}=Q_{i}P_{1}$
($i=a,b,c,d$; see (\ref{eq:isolegs})) are rays passing through the origin (dashed
lines in Figures \ref{fig:plantspace_vs_A1} and \ref{fig:plantspace_vs_cc}). Inequality
(\ref{eq:pollinator_ess_condition}) implies that the slopes of isolegs satisfy $Q_{d}<Q_{c}<Q_{b}<Q_{a}$
and, consequently, regions I, II, III, IV and V are ordered in a clockwise sequence
(Figure \ref{fig:plantspace_vs_A1}). As a result of this partition of the positive
quadrant, plant isoclines are defined piece-wise, and they are considerably more
complex when compared to the situation where pollinators have fixed preferences (cf.
Figure \ref{fig:plantspace_vs_A1} vs. Figure \ref{fig:nullclines}). Plant isoclines
in regions I, III, and V are easy to describe analytically (Appendix \ref{sec:appP1P2regions}).
However, in regions II and IV, plant isoclines are highly non-linear and although
they can be calculated using some computer algebra software (e.g., Mathematica),
the resulting expressions are too complex and they are not useful for further mathematical
analysis.

In what follows we will assume that each plant monoculture is viable, i.e., for P1

\begin{subequations}

\begin{equation}
\frac{a_{1}(r_{11}b_{11}A_{1}+r_{12}b_{12}A_{2})}{w_{1}+b_{11}A_{1}+b_{12}A_{2}}>m_{1},\label{viab1}
\end{equation}

\noindent and for P2

\begin{equation}
\frac{a_{2}(r_{21}b_{21}A_{1}+r_{22}b_{22}A_{2})}{w_{2}+b_{21}A_{1}+b_{22}A_{2}}>m_{2}.\label{viab2}
\end{equation}

\end{subequations}

This means that each plant equilibrates with pollinator densities when alone (section
\ref{plant_coexistence}). Then plant isoclines have the following general properties:
\begin{enumerate}
\item Isoclines consist of four connected segments, as shown by e.g., Figure \ref{fig:plantspace_vs_A1}a.
The isocline of plant P1 (P2) intersects the $P_{1}$ $(P_{2})$ axis at the origin
and at its pollinator-dependent carrying capacity in region V (I). These boundary
equilibria

\begin{subequations}

\begin{equation}
(P_{1},P_{2})=\bigg(K_{1}\left(1-\frac{m_{1}(w_{1}+b_{11}A_{1}+b_{12}A_{2})}{a_{1}(r_{11}b_{11}A_{1}+r_{12}b_{12}A_{2})}\right),0\bigg)\label{eq:P1boundary}
\end{equation}

\noindent and

\begin{equation}
(P_{1},P_{2})=\bigg(0,K_{2}\left(1-\frac{m_{2}(w_{2}+b_{21}A_{1}+b_{22}A_{2})}{a_{2}(r_{12}b_{21}A_{1}+r_{22}b_{22}A_{2})}\right)\bigg),\label{eq:P2boundary}
\end{equation}

\end{subequations}

\noindent are shown as filled circles on the axes of Figures \ref{fig:plantspace_vs_A1}
and \ref{fig:plantspace_vs_cc}. Appendix \ref{sec:appP1P2regions} shows that provided
these boundary equilibria exist (i.e., they are positive), they are locally asymptotically
stable.
\item The isoclines are linear in regions I, III and V, in which both pollinators are specialists.
Within these regions, $u_{1}$ and $v_{1}$ remain fixed at 0 or 1. If $c_{2}=0$
$(c_{1}=0)$ the isocline of plant P1 (P2) is vertical (horizontal), as shown in
Figure \ref{fig:plantspace_vs_A1} (cf., Figure \ref{fig:nullclines}a). If $c_{2}>0$
$(c_{1}>0)$ the isocline of plant P1 (P2) is negatively sloped within these regions,
as shown in Figure \ref{fig:plantspace_vs_cc} (cf., Figure \ref{fig:nullclines}b,c,d).
\item The isoclines are non-linear in regions II and IV, in which one pollinator is generalist
and the other specialist. The segment of the plant P1 (P2) isocline which is in region
II (IV) passes through the origin.
\item The isocline of plant P1 (P2) does not cross region I (V). This is because in region
I (V), plant P2 (P1) has two pollinators, but P1 (P2) has none and goes extinct in
this region.
\item The population density of plant P1 (P2) increases in the region below (to the left)
its isocline, and decreases in the region above (to the right).
\end{enumerate}
While there can be at most one interior plant equilibrium when pollinator preferences
for plants are fixed (section \ref{plant_coexistence}), there can be multiple interior
equilibria when preferences are adaptive, because isoclines intersect in multiple
points.

In the rest of this section we consider two particular scenarios that illustrate
the complexities of plant population dynamics under adaptive pollinator preferences:
\begin{itemize}
\item \emph{Scenario I: Plant population dynamics along the gradient in pollinator A1 density}.
In this scenario the density of pollinator A2 is kept fixed and both pollinators
are equally good for each plant $(r_{11}=r_{12},r_{21}=r_{22})$. Plants do not compete
for factors external to pollination $(c_{1}=c_{2}=0)$.
\item S\emph{cenario II: Plant population dynamics along the gradient in plant inter-specific
competition for external factors}. In this scenario we assume that plant inter-specific
competition is symmetric and we set $c=c_{1}=c_{2}$. We also assume that A1 (A2)
is the best pollinator of plant P1 (P2) $(r_{11}>r_{12}$, $r_{22}>r_{21})$.
\end{itemize}
Both scenarios are parameterized so that plant boundary equilibria (\ref{eq:P1boundary})
and (\ref{eq:P2boundary}) exist, i.e., pollinator densities are high enough so that
each plant can achieve a positive growth rate when alone.

The main purpose of scenario I is to explore how relative changes in pollinator densities
influence plant community composition. An important motivation is the growing interest
in the consequences of alien pollinator invasions \citep{traveset_richardson-tree06},
and the management of pollinator populations \citep{geslin_etal-aer17}. To focus
solely on plant competition for pollination services, we remove the effect of competition
for other factors (by setting competition coefficients equal to zero).

In Scenario II we explore how competition for external factors (e.g., space, nutrients)
influences competition between plants for pollinator preferences. Because of condition
(\ref{eq:pollinator_ess_condition}), this scenario also assumes that P1 (P2) and
A1 (A2) are better for one another. Such matching can be due to matching in plant
and pollinator morphologies \citep{fontaine_etal-plosbiol05}.

\subsubsection{Scenario I. Effects of changes in pollinator composition: Alternative plant stable
states}

\begin{figure}
\begin{centering}
\includegraphics[width=1\textwidth]{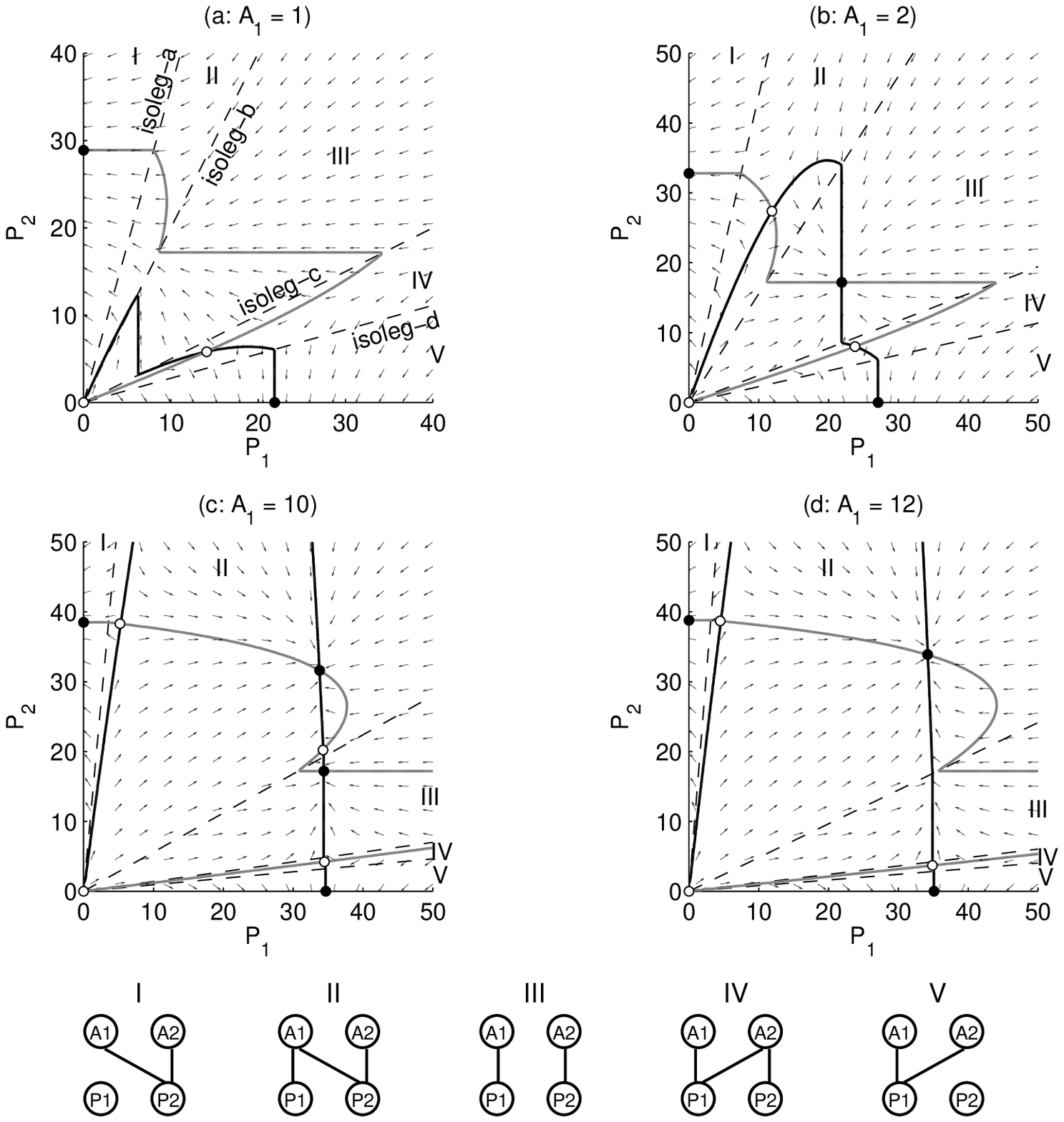} 
\par\end{centering}
\caption{\label{fig:plantspace_vs_A1}Isoclines of plants P1 (black) and P2 (gray), isolegs
(dashed lines), and vector field of plant population dynamics (arrows), under adaptive
pollinator preferences and increasing pollinator A1 density ($A_{1}$, scenario I).
Filled (open) circles represent stable (unstable) equilibria. Regions of pollinator
preference are defined in Table \ref{tab:regions_plaspace}, and corresponding interaction
topologies are indicated at the bottom. Parameters: $r_{ij}=0.1$, $m_{1}=0.01$,
$m_{2}=0.0075$, $c_{i}=0$, $a_{i}=0.4$, $w_{i}=0.25$, $b_{ij}=0.1$, $e_{11}=e_{22}=0.2$,
$e_{21}=e_{12}=0.1$, $K_{i}=50$, $A_{2}=1$. Note: parts of the isoclines are not
shown in (c,d), but these parts do not intersect at any equilibrium.}
\end{figure}

Figure \ref{fig:plantspace_vs_A1} illustrates plant population dynamics for scenario
I. Panel (a) shows the situation where pollinator A1 density is the same as pollinator
A2 density. Plant isoclines intersect in region IV, and the vector field indicates
that the corresponding equilibrium is unstable. Thus, there is bi-stability: depending
on initial conditions either plant P1 or P2 is excluded, and the plant community
becomes a monoculture. As density of pollinator A1 increases (panel b), the single
plant equilibria (\ref{eq:P1boundary}) and (\ref{eq:P2boundary}) increase too.
As a result, there are three isocline intersections in regions II, III and IV. The
equilibrium in region III is stable (because (\ref{eq:plant_stability}) holds, see
Appendix \ref{sec:appP1P2regions}) and the equilibria in regions II and IV are unstable.
Again, plant coexistence depends on initial conditions: if one plant is initially
too rare plant population dynamics will converge to a monoculture of the other plant,
but if the two plants are initially abundant enough, stable coexistence follows.
At the coexistence equilibrium pollinators specialize on different plants (see Table
\ref{tab:regions_plaspace}). In panel (c) pollinator A1 is more abundant than pollinator
A2, and two additional equilibria occur in region II, one stable and the other unstable.
Thus, there are two stable coexistence equilibria now (one in region II and the other
in region III). At the stable equilibrium that is in region II, pollinator A1 is
a generalist and A2 is a plant P2 specialist. As in panel (b), at the equilibrium
that lies in region III, pollinators specialize on different plants. Finally, in
panel (d), further increase in pollinator A1 leads to a single coexistence equilibrium
in region II where A1 is a generalist and A2 plant P2 specialist. 

\begin{figure}
\begin{centering}
\includegraphics[angle=-90,width=1\textwidth]{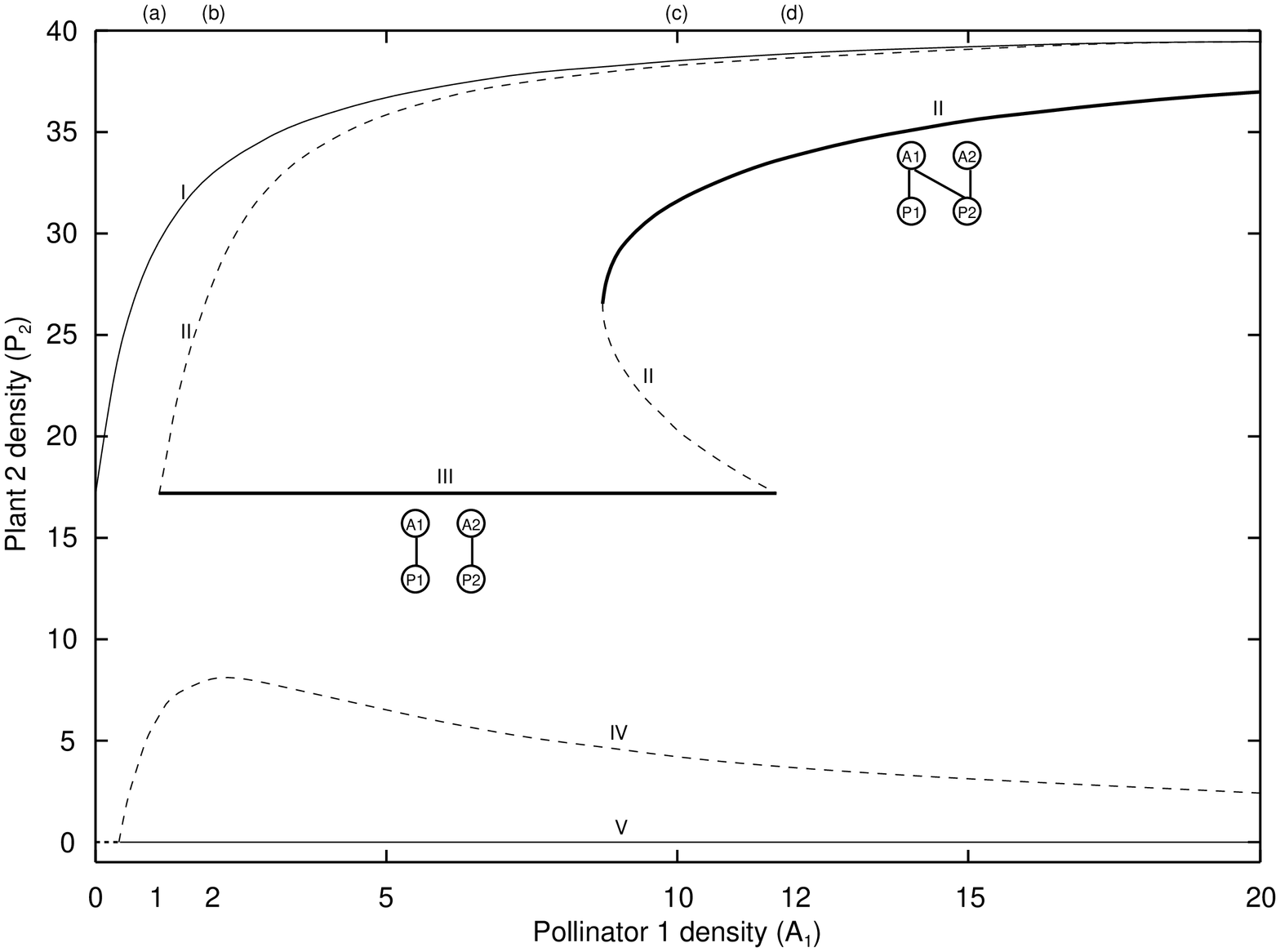} 
\par\end{centering}
\caption{\label{fig:bifA1}Bifurcation plot for plant P2 in scenario I. Thin solid lines represent
stable equilibria with one plant extinct. Thick solid lines represent stable coexistence
equilibria, next to corresponding interaction topology. Dashed lines represent unstable
equilibria. Roman numerals (I to V) indicate the location of equilibria within preference
regions given by the ESS (Table \ref{tab:regions_plaspace}). Labels along the top
of the plot correspond to panels in Figure \ref{fig:plantspace_vs_A1}.}
\end{figure}

Overall, the main effect of increasing pollinator A1 density with respect to A2,
is the reduction of region III where both pollinators specialize on different plants,
in favor of region II where A1 is a generalist and A2 a specialist. Here we see (Figure
\ref{fig:bifA1}) that along the gradient in $A_{1}$ density, the topology of the
interaction web changes. When population density of A1 is low, both pollinators specialize
on different plants. As population density of A1 increases, A1 becomes a generalist.
We also observe that plant P2 experiences \emph{hysteresis}: the stable equilibrium
in region III jumps to the stable equilibrium in region II at $A_{1}\approx11.7$
as pollinator density $A_{1}$ increases, but the stable equilibrium moving along
branch II jumps back to the stable equilibrium moving along branch III at $A_{1}\approx8.7$
when pollinator density $A_{1}$ decreases. Another important consequence of pollinator
A1 increase is that region I (V), in which P1 (P2) always decreases, become smaller.
This makes easier for plants to invade one another and achieve coexistence.

In summary, scenario I shows that: (i) adaptive foraging preferences can lead to
alternative plant coexistence stable states and (ii) continuous changes in pollinator
composition (i.e., $A_{1}:A_{2}$ ratio) produce discontinuous changes in plant\textendash pollinator
interaction structure.

\subsubsection{Scenario II. Effects of plant competition for external factors: Trait-mediated apparent
facilitation}

\begin{figure}
\begin{centering}
\includegraphics[width=1\textwidth]{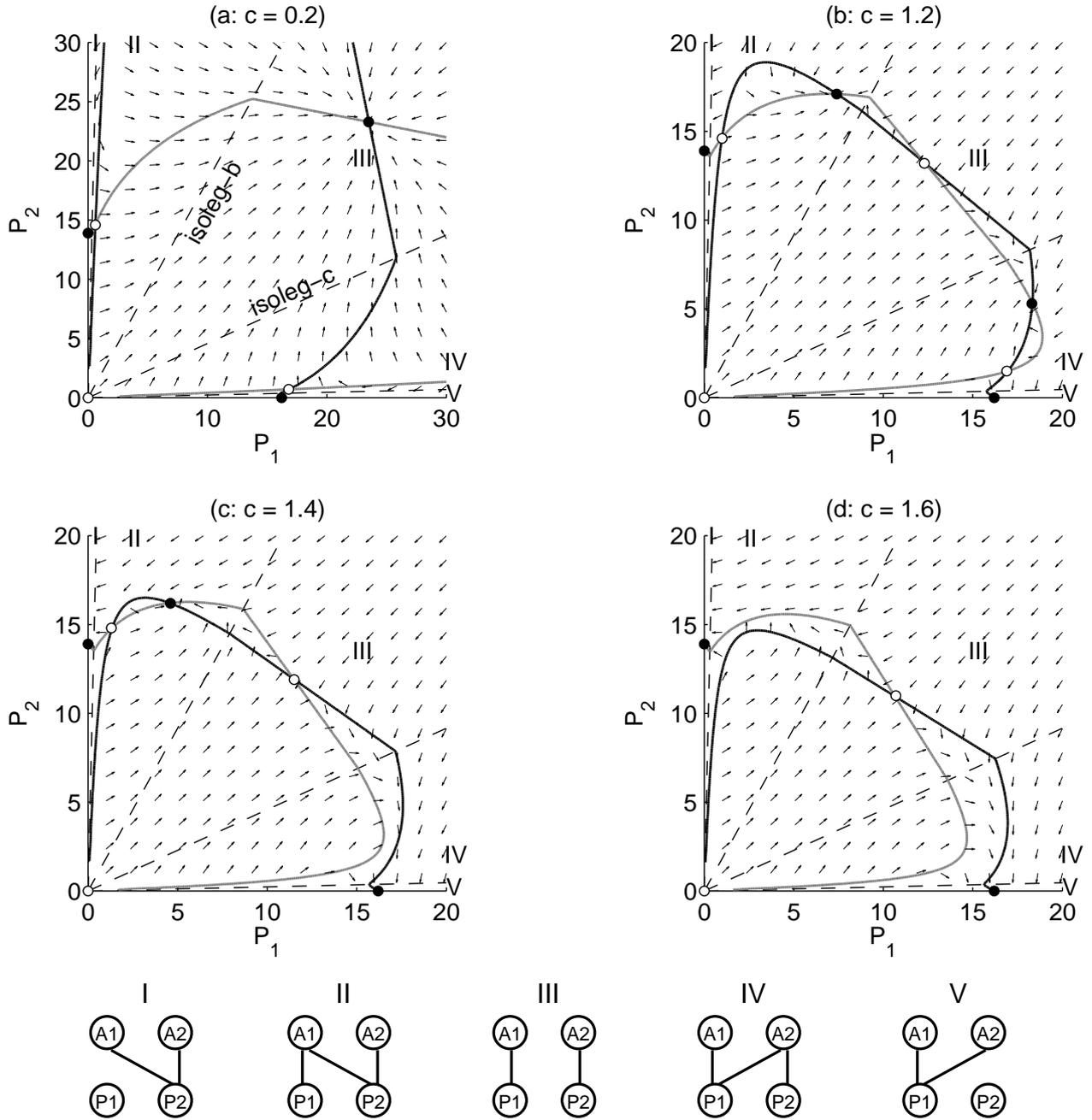} 
\par\end{centering}
\caption{\label{fig:plantspace_vs_cc}Isoclines of plants P1 (black) and P2 (gray), isolegs
(dashed lines), and vector field of plant population dynamics (arrows), under adaptive
pollinator preferences and with increasing plant competition ($c_{i}$, scenario
II). Filled (open) circles represent stable (unstable) equilibria. Regions of pollinator
preference are defined in Table \ref{tab:regions_plaspace}, and corresponding interaction
topologies are indicated at the bottom. Parameters: $r_{11}=r_{22}=0.5$, $r_{12}=r_{21}=0.1$,
$m_{i}=0.02$, $a_{i}=0.1$, $w_{i}=0.1$, $b_{ij}=0.1$, $e_{11}=e_{22}=0.2$, $e_{21}=e_{12}=0.1$,
$K_{i}=50$, $A_{1}=11$, $A_{2}=10$. Note: parts of the isoclines are not shown
in (a), but these parts do not intersect at any equilibrium.}
\end{figure}

Plant dynamics for scenario II are illustrated in Figure \ref{fig:plantspace_vs_cc}.
The isolegs (dashed lines, (\ref{eq:isolegs})) and boundary equilibria (\ref{eq:P1boundary})
and (\ref{eq:P2boundary}) do not change across panels (a\textendash d), because
they are independent of the competition coefficient $c=c_{1}=c_{2}$. Within regions
I, III and V the isoclines are linear while in regions II and IV they are non-linear.

When plant inter-specific competition is low (Figure \ref{fig:plantspace_vs_cc}a),
plant population dynamics are qualitatively similar to panels (b,d) in Figure \ref{fig:plantspace_vs_A1}
of scenario I, i.e., plants can coexist at a stable equilibrium. However, there is
an important qualitative difference here: At the coexistence equilibrium both plants
attain higher density when compared with their monoculture densities (boundary equilibria).
In other words, when inter-specific plant competition is weak, we observe mutual
plant facilitation. Let us consider the plant P1 boundary equilibrium in region V.
In this region P1 is pollinated by both pollinators. However, when A2 is a poor pollinator
for P1 (i.e., $r_{11}>r_{12}$ as assumed in Figure \ref{fig:plantspace_vs_cc}),
P1 can achieve a higher birth rate when it is pollinated by A1 only. So, if there
is an invasion of plant P2 from outside which moves the plant densities in region
III, pollinator A1 specializes on plant P1 and plant P2 is pollinated by its best
pollinator A2 only. Consequently, the P1 population equilibrium increases above its
monoculture level. Appendix \ref{sec:appP1P2regions} shows that the necessary condition
for this facilitation of plant P1 by the presence of P2 to happen is that $r_{11}/r_{12}>1+w_{1}/(b_{11}A_{1})$,
which means that pollinator A1 density must be high enough. In addition, such a facilitation
can happen only when inter-specific competition between plants is not too high. We
remark that this facilitation is not the usual one \citep{revilla_krivan-plosone16}
where an increase in one plant density increases the pollinator density which, in
turn, increases the other plant density. This mechanism cannot operate in the current
model that assumes pollinator population densities are fixed. The facilitation that
we observe here is due to changes in pollinator preferences, where by increasing
plant P2 density, pollinator A2 switches from pollinating plant P1 to pollinating
P2, which leads to an increase of P1 population density. To distinguish this mechanism
from density mediated facilitation caused by increase in pollinator density, we call
this mechanism indirect trait-mediated facilitation \citep[sensu][]{bolker_etal-ecology03}.

As inter-specific competition increases, plant equilibrium population densities in
region III will be decreasing below those they achieve in a monoculture (boundary
equilibria). When plant inter-specific competition is strong so that $c>1$, the
equilibrium in region III becomes unstable (i.e., (\ref{eq:plant_stability}) does
not hold, see also Appendix \ref{sec:appP1P2regions}), but plants can still coexist
at alternative stable states. In Figure \ref{fig:plantspace_vs_cc}b, the local dynamics
around the unstable equilibrium in region III is like in Figure \ref{fig:nullclines}c,
where perturbations cause either plant P1 to displace P2 or vice versa. Like in scenario
I, we have two alternative stable states at which both plants coexist. The most abundant
plant in each state is the one pollinated by both pollinators. Further increase of
the competition coefficient eliminates all equilibria in region IV, but the stable
equilibrium in region II remains, with pollinator A1 a generalist and A2 specialized
on P2 (Figure \ref{fig:plantspace_vs_cc}c). Finally, if competition is too strong
there are no equilibria in regions II and IV and we have mutual exclusion (Figure
\ref{fig:plantspace_vs_cc}d) where, depending on the initial conditions, one plant
outcompetes the other plant (cf. Figure \ref{fig:nullclines}c).

\begin{figure}
\begin{centering}
\includegraphics[angle=-90,width=1\textwidth]{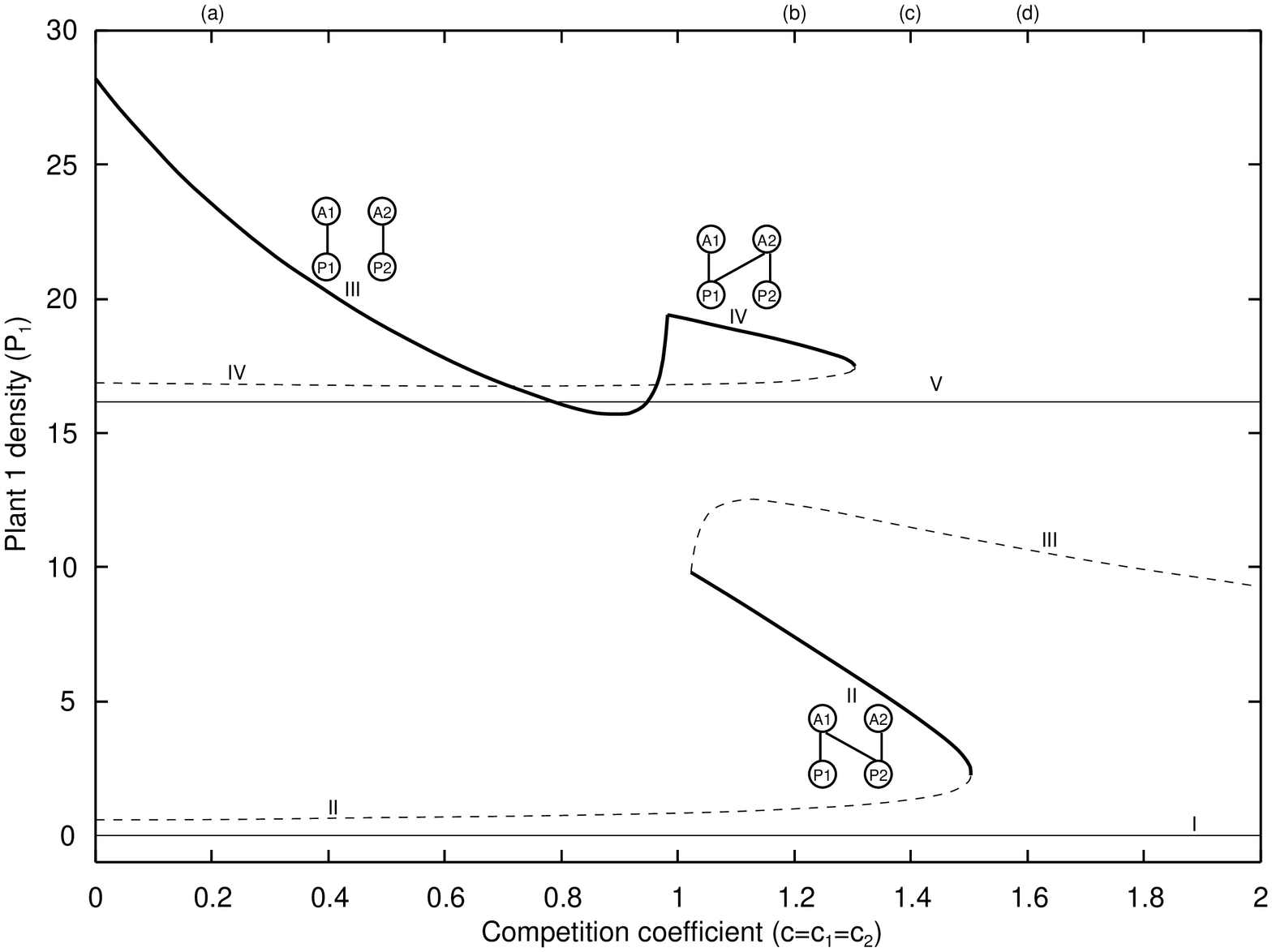} 
\par\end{centering}
\caption{\label{fig:bifCC}Bifurcation plot for plant P1 in scenario II. Lines and labels
follow the same conventions of Figure \ref{fig:bifA1}. Labels along the top of the
plot correspond to panels in Figure \ref{fig:plantspace_vs_cc}.}
\end{figure}

Figure \ref{fig:bifCC} shows the corresponding bifurcation plot for scenario II.
As competition for extrinsic factors (i.e., not for pollination) gets stronger, both
plant equilibrium densities tend to decrease, even in the region of alternative stable
states $(1\apprle c\apprle1.3)$ where P1 can be either abundant (stable IV branch)
or rare (stable II branch). There is only a small region where plant P1 increases
with competition $(0.9\apprle c\apprle1)$, i.e., where the combined effects of exploitative
competition and competition for pollination (i.e., trait-mediated plant facilitation)
is more favorable for P1 than for P2 (which decreases, not shown). Notice that in
comparison to Figure \ref{fig:bifA1} which shows transitions between two stable
interaction topologies, Figure \ref{fig:bifCC} shows transitions between three stable
interaction topologies.

In summary, scenario II shows that: (i) adaptive foraging preferences can result
in indirect trait-mediated plant\textendash plant facilitation, by matching plants
with their best pollinators; (ii) continuous changes in competition for factors external
to pollination can produce discontinuous changes in interaction structure and coexistence
for plants competing for pollination services; and (iii) plants can coexist even
when inter-specific competition is stronger than intra-specific competition for factors
other than pollination. In the next section we use isolegs and isoclines to study
pollinator\textendash pollinator competition.

\subsection{Pollinators compete for plant resources\label{subsec:polli_comp_adpref}}

In this section we analyze population dynamics of adaptive pollinators at fixed plant
densities. Unlike in the case of fixed preferences (Figure \ref{fig:nullclines_pollinators}),
now we must partition the first quadrant of the pollinator plane $A_{1}A_{2}$ into
different regions using isolegs (Figure \ref{fig:phase_space_pollinators}), according
to Table \ref{tab:regions_polspace} (see Appendix \ref{sec:appA1A2coexflex}). The
isolegs are linear in $A_{1}$ and they are given by $A_{2}=S_{i}(P_{1},P_{2})A_{1}+I_{i}(P_{1},P_{2})$
(where $i=a,b,c,d$) where slopes and intercepts are

\begin{equation}
\begin{array}{lc}
S_{a}(P_{1},P_{2})=-\frac{b_{21}}{b_{22}}, & I_{a}(P_{1},P_{2})=\frac{a_{2}b_{21}e_{21}P_{2}w_{1}-a_{1}b_{11}e_{11}P_{1}w_{2}}{a_{1}b_{11}b_{22}e_{11}P_{1}},\\[0.5cm]
S_{b}(P_{1},P_{2})=\frac{a_{2}e_{21}b_{21}P_{2}}{a_{1}e_{11}b_{22}P_{1}}, & I_{b}(P_{1},P_{2})=\frac{a_{2}b_{21}e_{21}P_{2}w_{1}-a_{1}b_{11}e_{11}P_{1}w_{2}}{a_{1}b_{11}b_{22}e_{11}P_{1}},\\[0.5cm]
S_{c}(P_{1},P_{2})=\frac{a_{2}e_{22}b_{11}P_{2}}{a_{1}e_{12}b_{12}P_{1}}, & I_{c}(P_{1},P_{2})=\frac{a_{2}b_{22}e_{22}P_{2}w_{1}-a_{1}b_{12}e_{12}P_{1}w_{2}}{a_{1}b_{12}b_{22}e_{12}P_{1}},\\[0.5cm]
S_{d}(P_{1},P_{2})=-\frac{b_{11}}{b_{12}}, & I_{d}(P_{1},P_{2})=\frac{a_{1}b_{12}e_{12}P_{1}w_{2}-a_{2}b_{22}e_{22}P_{2}w_{1}}{a_{2}b_{22}b_{12}e_{22}P_{2}}.
\end{array}\label{eq:slopes_intercepts}
\end{equation}

\begin{table}
\begin{tabular}{cccl}
\hline 
Region  & Conditions  & ESS $(u_{1},v_{1})$  & Description\tabularnewline
\hline 
I  & $A_{2}<S_{a}A_{1}+I_{a}$  & $(0,0)$  & A1 \& A2 specialize on P2\tabularnewline
II  & $S_{a}A_{1}+I_{a}<A_{2}<S_{b}A_{1}+I_{b}$  & $(u_{1}^{*},0)$  & A1 generalist, A2 specializes on P2\tabularnewline
III  & $S_{b}A_{1}+I_{b}<A_{2}<S_{c}A_{1}+I_{c}$  & $(1,0)$  & A1 specializes on P1, A2 specializes on P2\tabularnewline
IV  & $\max\{S_{c}A_{1}+I_{c}\,,\,S_{d}A_{1}+I_{d}\}<A_{2}$  & $(1,v_{1}^{*})$  & A1 specializes on P1, A2 generalist\tabularnewline
V  & $A_{2}<S_{d}A_{1}+I_{d}$  & $(1,1)$  & A1 \& A2 specialize on P1\tabularnewline
\hline 
\end{tabular}\caption{\label{tab:regions_polspace}ESS as a function of pollinator densities. Isoleg slopes
$S_{i}(P_{1},P_{2})$ and intercepts with the $A_{2}$ axis $I_{i}(P_{1},P_{2})$
$(i=a,b,c,d)$ are given by (\ref{eq:slopes_intercepts}), and $u_{1}^{*}$ and $v_{1}^{*}$
by (\ref{eq:ess}).}
\end{table}

Compared to isolegs in the plant plane (Figures \ref{fig:plantspace_vs_A1} and \ref{fig:plantspace_vs_cc}),
in the pollinator plane isolegs neither pass through the origin, nor all have positive
slopes. Thus, for given parameter values and plant population densities not all regions
from Table \ref{tab:regions_polspace} exist in the positive quadrant. In general:
\begin{enumerate}
\item Regions II, III and IV always occur (see Figure \ref{fig:phase_space_pollinators}).
They are separated by the isoleg-b ($A_{2}=S_{b}A_{1}+I_{b}$) and the isoleg-c ($A_{2}=S_{c}A_{1}+I_{c}$)
with positive slopes $S_{b}$ and $S_{c}$, respectively. These isolegs do not intersect
in the first quadrant of the $A_{1}A_{2}$ plane (Appendix \ref{sec:appA1A2coexflex}).
\item Because of (\ref{eq:pollinator_ess_condition}) the isoleg-c separating IV and III
is steeper than the isoleg-b separating III and II $(S_{c}>S_{b})$. Thus, regions
II, III and IV are ordered in a counter-clockwise sequence in the positive $A_{1}A_{2}$
plane.
\item Regions I and V are separated from regions II and IV, respectively, by isoleg-a and
isoleg-d with negative slopes $S_{a}$ and $S_{d}$. Appendix \ref{sec:appA1A2coexflex}
shows that at most one of these two regions can exist for given parameters and plant
population densities. E.g., in Figure \ref{fig:phase_space_pollinators}a neither
of the two regions exist, while in \ref{fig:phase_space_pollinators}b region I exists.
\end{enumerate}
\begin{figure}
\begin{centering}
\includegraphics[width=1\textwidth]{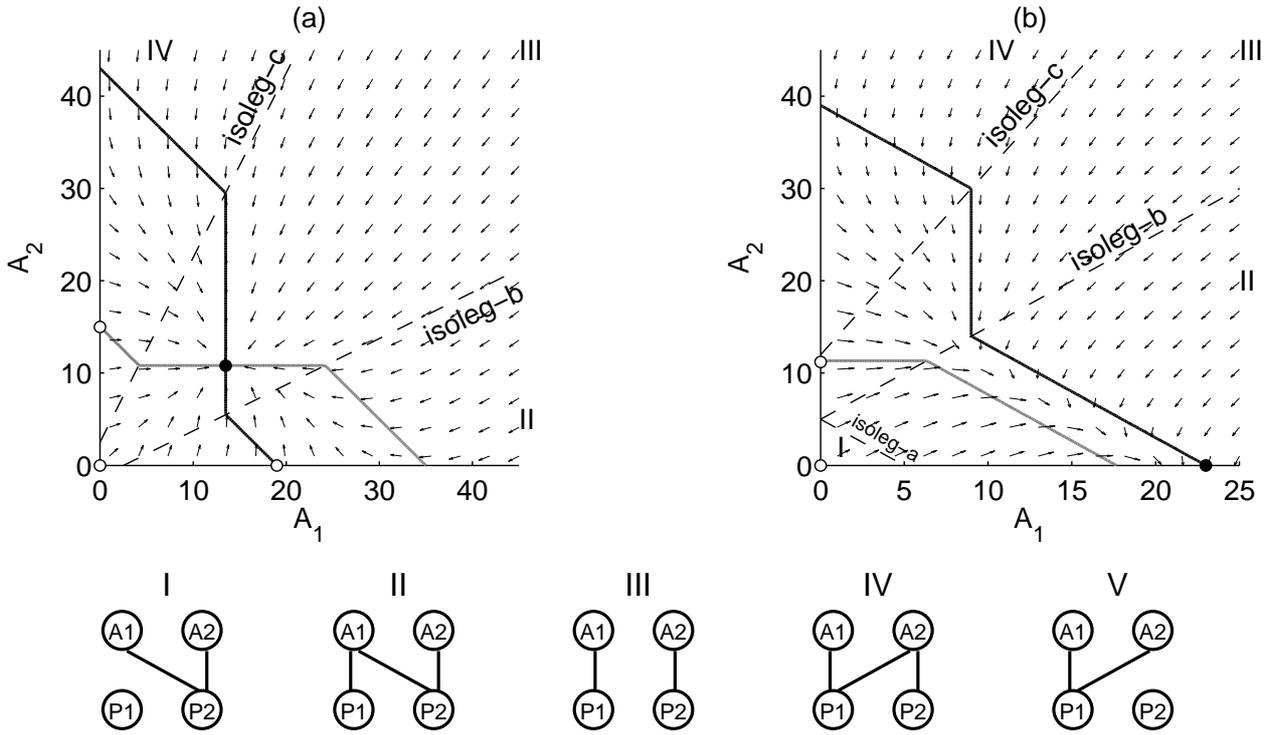} 
\par\end{centering}
\caption{\label{fig:phase_space_pollinators}Isoclines of pollinators A1 (black) and A2 (gray),
isolegs (dashed lines), and dynamics (vector field), under adaptive pollinator preferences.
Filled (open) circles represent stable (unstable) equilibria. Regions of pollinator
preference are defined in Table \ref{tab:regions_polspace}, and corresponding interaction
topologies are indicated at the bottom. Parameters: $a_{i}=0.4$, $b_{ij}=0.1$,
$d_{1}=0.1$, $d_{2}=0.12$, $P_{1}=P_{2}=20$ in all panels; (a) $e_{11}=e_{22}=0.2$,
$e_{12}=e_{21}=0.1$, $w_{i}=0.1$; (b) $e_{11}=e_{21}=e_{12}=0.2$, $e_{12}=0.1$,
$w_{1}=0.7,w_{2}=0.2$.}
\end{figure}

The partition of the pollinator plane results in pollinator isoclines that are more
complex than in the case of fixed preferences, but considerably simpler than plant
isoclines in section \ref{subsec:plant_comp_adpref}. The isoclines consist of three
(e.g., Figure \ref{fig:phase_space_pollinators}a) or two connected segments (e.g.,
the pollinator A2 isocline in Figure \ref{fig:phase_space_pollinators}b). Regions
I and V contain no isocline segments. The segments within regions II and IV are linearly
decreasing, and both isoclines are parallel in these two regions (see Appendix \ref{sec:appA1A2coexflex}).
Thus, generically, pollinators cannot coexist within regions II or IV. This is unlike
the case with fixed preferences, where the specialist has a linear isocline and the
generalist a curved isocline (Figure \ref{fig:nullclines_pollinators}e,f). Finally,
the segments of isoclines in region III are vertical for A1 and horizontal for A2,
because pollinators specialize on different plants (like in Figure \ref{fig:nullclines_pollinators}a).
Thus, pollinator coexistence can only occur in region III when the vertical segment
of A1 and the horizontal segment of A2 intersect, as shown in Figure \ref{fig:phase_space_pollinators}a.
Given (\ref{eq:pollinator_ess_condition}), Appendix \ref{sec:appA1A2coexflex} demonstrates
that pollinator coexistence by mutual invasion requires

\begin{equation}
\frac{b_{21}e_{21}}{b_{22}e_{22}}<\frac{d_{1}}{d_{2}}<\frac{b_{11}e_{11}}{b_{12}e_{12}},\label{eq:A1A2mutinv}
\end{equation}

\noindent leading to a stable equilibrium in region III. If $d_{1}/d_{2}$ is too
low to meet above inequalities, pollinator A2 goes extinct as shown in Figure \ref{fig:phase_space_pollinators}b,
and if $d_{1}/d_{2}$ is too large A1 goes extinct instead. The coexistence scenario
in Figure \ref{fig:phase_space_pollinators}a is called the \emph{ghost of competition
past} \citep{connell-oikos80}, because competition between pollinators causes selection
for different plants which ends competition in the long term. What happens here is
that the preference trade-off ( $u_{1}+u_{2}=1$ and $v_{1}+v_{2}=1$) causes disadvantage
for the generalist when combining its best and worst resources. This is not the case
for the specialist that fully commits to its best resource. Thus, in region II (IV),
selection drives A1 (A2) individuals to increase preference towards its preferred
plant P1 (P2). As a consequence, pollinators specialize on different plants.

In summary, the results show that population dynamics of two adaptable pollinators
competing for two plants do not allow stable coexistence between two generalists,
one generalist and one specialist, and two specialists on the same plant. In other
words, coexistence demands absolute niche segregation where each pollinator has its
own plant.

\section{Discussion\label{sec:Discussion}}

In this article we study how pollinator adaptation affects coexistence in a community
module consisting of two plants and two pollinators. We assume that pollinators preferences
for plants are adaptive and they correspond to evolutionarily stable strategies (ESS)
at given plant and pollinator densities. Such strategies cannot be invaded by any
other mutants with different strategies. We prove that the strategy where both pollinators
are generalists is never evolutionarily stable. Then we study plant\textendash plant
and pollinator\textendash pollinator population dynamics. We observe that at fixed
pollinator densities, adaptive pollinator preferences for plants lead to complex
plant dynamics characterized by alternative stable states. Such alternative states
do not exist when interaction strengths between pollinators and plants are fixed.
We also observe a trait-mediated facilitation \citep[sensu][]{bolker_etal-ecology03}
between plants due to changes in pollinator preferences where introduction of an
alternative plant can increase population density of the original plant, without
increasing pollinator density. When plant densities are fixed, our analysis of pollinator\textendash only
dynamics shows that a stable coexistence of a generalist and a specialist pollinator
is not possible when both pollinators are adaptive foragers. Thus, at the pollinator
coexistence equilibrium, each plant must have its own pollinator.

Our analyses combine an evolutionary approach with population dynamics. The evolutionary
approach 
is based on isolegs \citep{rosenzweig-ecology81,pimm_rosenzweig-oikos81,krivan_sirot-amnat02}
analysis. Isolegs split the plant (or pollinator) phase space into several regions
that are characterized by pollinator specialization or generalism. The population
dynamic approach is based on isocline analysis. When compared to standard models
of population dynamics, the case where pollinators are adaptive foragers leads to
isoclines that are defined piece-wise depending on the pollinator optimal strategy.
For example, when interaction strengths between pollinators and plants are fixed
(i.e., pollinators are inflexible foragers), plant\textendash plant dynamics follow
the Lotka\textendash Volterra competition model with isoclines being straight lines
(Figure \ref{fig:nullclines}, top row). However, when pollinators are adaptive foragers,
plant isoclines are highly non-linear (e.g., Figure \ref{fig:plantspace_vs_A1}).
It is this emerging non-linearity that shows striking consequences of adaptive pollinator
behavior in the interaction web studied in this article.

In order to get insights on plant and pollinator coexistence, we assume that one
mutualistic guild, the plants or the pollinators, stays at constant densities, while
the other undergoes population dynamics. This is a limitation, but such conditions
are not uncommon in nature. E.g., plants can be long lived trees or shrubs, while
pollinators can be comparatively short lived, e.g., insects. The assumption of pollinator
densities being constant while plants undergo population dynamics can represent situations
where plants are short lived (e.g., grasses or forbs), while pollinator densities
are mainly controlled by factors other than mutualism (e.g., pollinators may be limited
by availability of artificial beehives or tree holes). Another possibility is that
plant dynamics take place in a small locality or a patch, and this patch has a certain
pollinator carrying capacity which is rapidly filled by visiting pollinators \citep{feldman_etal-oikos04}
coming from a much larger region. This can be the case of massively introduced managed
pollinators, spilling over from mass flowering crops into wild plant communities
\citep{geslin_etal-aer17}.

\subsection{Adaptive pollinator preferences}

When two pollinators compete for resources provided by two plants, we predict five
qualitatively different pollinator preferences that are evolutionarily stable (Table
\ref{tab:regions_plaspace}). These strategies are characterized either as full specialization
of a pollinator on a single plant or generalism. We proved that the situation where
both pollinators are generalists is never evolutionarily stable and it should not
be observed in nature. The distribution of pollinator preferences is similar to the
ideal free distribution (IFD) of two consumers using two resource patches \citep{krivan-tpb03}.

Pollinator preferences were derived under conditions of low species diversity (only
four species), and constant population densities. Interestingly, such conditions
are approximated in the experiments of \citet{fontaine_etal-plosbiol05}. These authors
used two plant groups: plants with open (P1), and tubular (P2) flowers; and two pollinator
groups: syrphid flies (A1), and bumblebees (A2). Each group consisted of three species.
This diversity ensures that each pollinator group can use each plant group. However,
syrphid flies are morphologically better adapted to open flowers, whereas bumblebees
are better adapted to tubular flowers. Plants and pollinators interacted at fixed
densities within cages. One experiment found that when alone, each pollinator group
displayed generalism. However, when together, syrphids tended to visit open flowers
almost exclusively, whereas bumblebees tended to maintain their generalism. This
observation corresponds with our partially mixed ESS with one specialist and one
generalist pollinator. Further experimentation, with controlled variation of P1:P2
and A1:A2 abundance ratios, will be necessary to test our predictions (Table \ref{tab:regions_plaspace}).

\subsection{How adaptive preferences change plant coexistence}

Analysis of plant dynamics when pollinator densities are fixed indicates that pollinator
preferences can modify the plant community to a large extent. Under fixed pollinator
preferences, plant population dynamics are described by the Lotka\textendash Volterra
competition model. Thus, plants either coexist at an equilibrium, or one plant is
outcompeted by the other plant (Figure \ref{fig:nullclines}). In the bi-stable case
when initial conditions determine the outcome of competition (Figure \ref{fig:nullclines}c),
the preferred plant that survives has a larger domain of attraction so it is expected
to win more frequently. When pollinator preferences are adaptive, initial conditions
have major effects on plant coexistence for three main reasons. First, since pollination
is obligatory for both plants, coexistence requires that no plant is initially too
rare, because otherwise positive feedbacks make the rare plant less preferred and
the common plant more preferred (\emph{the rich get richer and the poor get poorer}
situation), causing the rare plant extinction. The same feedbacks prevent invasion
of rare plants, unless invaders start above minimum density thresholds. Second, pollinator
adaptation enables alternative stable states in plant coexistence. Third, plants
can coexist even when their inter-specific competition is so strong that one plant
would be outcompeted when pollinators were inflexible foragers.

Many mutualistic models predict critical transitions in community composition as
a result of an environmental stress (e.g., warming, habitat fragmentation, changes
in phenology). These critical transitions can lead to states of very low diversity,
or community collapse when mutualism is obligatory. In large communities, critical
transitions are preceded by a gradual accumulation of species extinctions that cause
interaction loss (e.g., simulated by random species removal, \citealp{jellelever_etal-ecolett14}).
On a much smaller scale (only four species) our scenario I, where the density of
pollinator A1 increases while the density of the second pollinator A2 is kept fixed,
demonstrates critical transitions (i.e., discontinuous changes both in numbers and
the interaction topology) due to interaction loss. In this scenario, transitions
between single and alternative stable states in the plant community are due to switches
in one pollinator (A1) strategy. When the pollinator is rare it specializes on the
best plant (Fig. \ref{fig:plantspace_vs_A1}b). As its population increases the pollinator
switches to a generalist (Fig. \ref{fig:plantspace_vs_A1}c), in response to increased
competition. We do not have empirical evidence for transitions like in scenario I,
but we can hypothesize one of practical importance. Consider a managed pollinator
(e.g., A1 = honeybees) coexisting with wild pollinators (e.g., A2 = bumblebees).
We assume that managed pollinators start with high densities e.g., thanks to artificial
beehives. Because of competition for plants this large population will generalize
\citep{fontaine_etal-joe08}, pollinating many plants and maintaining high plant
diversity (in Figure \ref{fig:bifA1} this corresponds to pollinator A1 above $A_{1}\approx8.7$
and plant P2 density given by the solid curve labeled by II). A parasite infestation
will cause the managed pollinator population to collapse to much lower densities
\citep{guzmannovoa_etal-apidologie10} (below $A_{1}\approx8.7$ in Figure \ref{fig:bifA1}).
Competition between pollinators for plants will be lower and they will specialize
(pollinator A1 specializes on P1 in Figure \ref{fig:bifA1}). This will lead to a
critical transition in the plant community where P2 density drops to $P_{2}\approx17$
(solid line labeled by III). In order to revert back to the condition where P2 had
a higher density ($P_{2}\approx27$ and larger, solid line labeled by II), pollinator
A1 must become generalist again, but due to hysteresis the density of this managed
pollinator must be raised to levels higher than before the collapse (i.e., A1 must
reach population density above $A_{1}\approx11.7$ in Figure \ref{fig:bifA1}), e.g.,
by providing additional beehives. This hypothetical scenario could be tested using
semi-closed experimental plant communities, by controlling the access of massively
introduced managed pollinators living nearby \citep{geslin_etal-aer17}.

Competition for pollinator preferences can result in plant coexistence at densities
that are smaller (scenario I, Figure \ref{fig:plantspace_vs_A1}, specially for P2)
or larger (scenario II, Figure \ref{fig:plantspace_vs_cc}a) than the densities when
each plant is alone. The first prediction was widely confirmed empirically \citep{chittka_schuerkens-nature01,aizen_etal-nwephytol14}.
Regarding the second prediction, the experiments of \citet{fontaine_etal-plosbiol05}
discussed before indicate that plant facilitation is a potentially realistic outcome.
In that experiment, plants with open flowers (P1) were better adapted to syrphid
flies (A1) and vice-versa, whereas plants with tubular flowers (P2) were better adapted
to bumblebees (A2). Bumblebees are generalists and they are slightly better at using
tubular flowers. When the four groups were placed together, competition forced syrphids
to concentrate on open flowers and bumblebees to prefer tubular flowers. At the end
of experiment each plant group was taken care of by its best pollinator group, and
ended up producing more seeds. This experiment and our predictions demonstrate that
given enough functional diversity, i.e., differences in plant and pollinator functional
traits, adaptive pollination can improve not only pollinator coexistence but also
plant coexistence to the point where plants can end up facilitating one another indirectly.
We note that this facilitation between plants can be due to changes in pollinator
densities (indirect density-mediated facilitation), or due to changes in trait (indirect
trait-mediated facilitation, \citealp{bolker_etal-ecology03}) which is caused by
changes in pollinator preferences for plants. The interplay between such indirect
effects with direct competition between plants for other factors (e.g., space or
nutrients, described by competition coefficients), can give rise to alternative stable
coexistence states (Figure \ref{fig:plantspace_vs_cc}b) \citep{hernandez-rspb98,gerla_mooij-tpb14,zhang_etal-ecomod15,holland_deangelis-ecolett09,holland_etal-amnat02}.

\subsection{How adaptive preferences change pollinator coexistence}

The analysis of pollinator population dynamics described by equations (\ref{eq:ode_a1},\ref{eq:ode_a2})
predicts that adaptation of pollinator preferences results in competitive outcomes
that are similar to those with fixed preferences: both pollinators can coexist, one
always excludes the other, or initial conditions determine which pollinator survives
and which goes extinct. In particular, there are no alternative stable states such
as we see in the plant sub-system. There are, however, important qualitative differences
in the community interaction topology. We already know that the case where both pollinators
are generalists is not evolutionarily stable and it cannot occur. However, pollinator
population dynamics also exclude pollinator stable coexistence in the case where
one pollinator is a specialist and the other a generalist. Thus, when pollinators
adapt their foraging preferences with changing population numbers, only pollinators
that specialize on different plants can coexist (Figure \ref{fig:phase_space_pollinators}a).
As a result, both pollinators stop to compete (the \emph{ghost of competition past,}
\citealp{connell-oikos80}).

We get similar conclusions from numerical simulations of the full four species system
(\ref{eq:ode_system}) with adaptive pollinator preferences (Table \ref{tab:regions_plaspace}
or \ref{tab:regions_polspace}): pollinators either specialize on different plants,
or specialist pollinators are excluded by generalists (Appendix \ref{sec:dynamics_module}
shows representative simulations). These results suggest that plant coexistence at
alternative states is unlikely when both plant and pollinator dynamics operate on
similar time scales.

These conclusions have important implications for systems containing many pollinator
species. Most real plant\textendash pollinator interaction networks are nested \citep{bascompte_jordano-arees07}.
This means that a minority of generalist pollinators can interact with many plants,
but a majority of more specialized pollinators interact with a few plants only, typically
subsets of the plants used by the generalists. This causes a disadvantage for specialized
pollinators that have to compete for resources with generalist competitors. Numerical
simulations show that adaptive foraging tends to reduce the effect of nestedness
on pollinator diet overlap \citep{valdovinos_etal-ecolett16}. As a consequence,
specialist pollinators experience less competition, pollination for plants with less
pollinators becomes more efficient, and more plants and pollinators can coexist in
the long term. We observe the same mechanism in our two-pollinator\textendash two-plant
interaction module. For example, consider a generalist pollinator A1 and a specialist
A2 (i.e., $0<u_{1}<1,v_{1}=0$) as a caricature of a nested network. Such interaction
topology can be dynamically stable when preferences of generalist pollinators are
fixed (Figure \ref{fig:nullclines_pollinators}e), but not when preferences adapt
in which case either (i) both pollinators specialize on different plants (Figure
\ref{fig:phase_space_pollinators}a) or (ii) the specialist goes extinct (Figure
\ref{fig:phase_space_pollinators}b). In the first case nestedness is eliminated
as the pollinator A1 becomes a specialist.

\subsection{Conclusions}

As the take-home-message, our analysis of a two-plant\textendash two-pollinator interaction
web demonstrates that adaptation of pollinator preferences for plants causes important
changes in the structure and dynamics of plant and pollinator communities. First,
when pollinator preferences are fixed, interactions between plants follow the Lotka\textendash Volterra
competitive dynamics when pollinator densities are held constant. When plant densities
are fixed, coexistence of generalist pollinators is possible. Second, when pollinator
preferences adapt in order to maximize fitness, plant competitive dynamics become
more complex and plant coexistence at alternative stable states and indirect plant\textendash plant
facilitation is possible, if pollinator densities are held constant. At fixed plant
densities competition between adaptive pollinators requires pollinators specialize
on different plants.

\section*{Acknowledgements}

We thank Francisco Encinas\textendash Viso and two anonymous reviewers for comments
and suggestions. Support provided by the Institute of Entomology (RVO:60077344) is
acknowledged. This project has received funding from the European Union's Horizon
2020 research and innovation programme under the Marie Sk\l{}odowska-Curie grant
agreement No 690817.

\bibliographystyle{chicago}
\bibliography{competition_for_pollination}

\appendix
\newpage{}

\renewcommand{\theequation}{A.\arabic{equation}}
\setcounter{equation}{0}
\renewcommand{\thefigure}{A.\arabic{figure}}
\setcounter{figure}{0}
\renewcommand{\thetable}{A.\arabic{table}}
\setcounter{table}{0}

\section{Coexistence conditions for pollinators with fixed preferences\label{sec:appA1A2coex}}

We set $\alpha_{ij}=a_{i}P_{i}e_{ij}$ and $\beta_{i1}=u_{i}b_{i1},$ $\beta_{i2}=v_{i}b_{i2}$
and re-write the pollinator sub-system (\ref{eq:ode_a1}, \ref{eq:ode_a2}) as

\begin{align}
\frac{dA_{1}}{dt} & =\left\{ \frac{\alpha_{11}\beta_{11}}{w_{1}+\beta_{11}A_{1}+\beta_{12}A_{2}}+\frac{\alpha_{21}\beta_{21}}{w_{2}+\beta_{21}A_{1}+\beta_{22}A_{2}}-d_{1}\right\} A_{1}\nonumber \\
\frac{dA_{2}}{dt} & =\left\{ \frac{\alpha_{12}\beta_{12}}{w_{1}+\beta_{11}A_{1}+\beta_{12}A_{2}}+\frac{\alpha_{22}\beta_{22}}{w_{2}+\beta_{21}A_{1}+\beta_{22}A_{2}}-d_{2}\right\} A_{2}.\label{eq:pol-subsystem}
\end{align}

Model (\ref{eq:pol-subsystem}) has the trivial equilibrium $(A_{1},A_{2})=(0,0).$
When $A_{2}=0$, the per-capita population growth rate $dA_{1}/(A_{1}dt)$ of pollinator
A1 decreases monotonically with $A_{1}$. Provided the per capita birth rate of pollinator
1 when $A_{1}=A_{2}=0$ is larger than is its per capita population death rate, i.e.,
\begin{equation}
\alpha_{11}\beta_{11}w_{2}+\alpha_{21}\beta_{21}w_{1}>d_{1}w_{1}w_{2}\label{eq:viable1}
\end{equation}
there is exactly one A1-only equilibrium

\[
(A_{1},0)=\bigg(\frac{-b+\sqrt{b^{2}-4ac}}{2a},0\bigg),
\]

\noindent where $a=d_{1}\beta_{11}\beta_{21}$, $b=d_{1}(w_{2}\beta_{11}+w_{1}\beta_{21})-(\alpha_{11}+\alpha_{21})\beta_{11}\beta_{21}$
and $c=d_{1}w_{1}w_{2}-w_{2}\alpha_{11}\beta_{11}-w_{1}\alpha_{21}\beta_{21}$.

If the opposite inequality in (\ref{eq:viable1}) holds, the per-capita population
growth rate of pollinator 1 is always negative and the pollinator goes extinct. By
symmetry, if

\begin{equation}
\alpha_{12}\beta_{12}w_{2}+\alpha_{22}\beta_{22}w_{1}>d_{2}w_{1}w_{2}\label{eq:viable2}
\end{equation}

\noindent there is a unique A2-only equilibrium.

Provided $\beta_{11}\beta_{22}-\beta_{12}\beta_{21}\ne0$, $d_{2}\alpha_{11}\beta_{11}-d_{1}\alpha_{12}\beta_{12}\ne0$,
and $d_{2}\alpha_{21}\beta_{21}-d_{1}\alpha_{22}\beta_{22}\ne0,$ model (\ref{eq:pol-subsystem})
has at most one coexistence equilibrium

\begin{align}
\hat{A}_{1} & =\frac{w_{2}\beta_{12}-w_{1}\beta_{22}+\frac{(\alpha_{11}\alpha_{22}\beta_{11}\beta_{22}-\alpha_{12}\alpha_{21}\beta_{12}\beta_{21})[d_{1}(\alpha_{12}+\alpha_{22})\beta_{12}\beta_{22}-d_{2}(\alpha_{21}\beta_{12}\beta_{21}+\alpha_{11}\beta_{11}\beta_{22})]}{(d_{2}\alpha_{11}\beta_{11}-d_{1}\alpha_{12}\beta_{12})(d_{2}\alpha_{21}\beta_{21}-d_{1}\alpha_{22}\beta_{22})}}{\beta_{11}\beta_{22}-\beta_{12}\beta_{21}}\nonumber \\
\hat{A}_{2} & =\frac{w_{1}\beta_{21}-w_{2}\beta_{11}+\frac{(\alpha_{11}\alpha_{22}\beta_{11}\beta_{22}-\alpha_{12}\alpha_{21}\beta_{12}\beta_{21})[d_{2}(\alpha_{11}+\alpha_{21})\beta_{11}\beta_{21}-d_{1}(\alpha_{12}\beta_{12}\beta_{21}+\alpha_{22}\beta_{11}\beta_{22})]}{(d_{2}\alpha_{11}\beta_{11}-d_{1}\alpha_{12}\beta_{12})(d_{2}\alpha_{21}\beta_{21}-d_{1}\alpha_{22}\beta_{22})}}{\beta_{11}\beta_{22}-\beta_{12}\beta_{21}}\label{eq:pol-equilibrium}
\end{align}

\noindent if $\hat{A}_{1}>0$ and $\hat{A}_{2}>0.$

Now we study the local asymptotic stability of the equilibria. The jacobian of (\ref{eq:pol-subsystem})
is

\begin{equation}
\mathbf{J}=\left[\begin{array}{cc}
G_{1}-A_{1}\left(\frac{\alpha_{11}\beta_{11}^{2}}{W_{1}^{2}}+\frac{\alpha_{21}\beta_{21}^{2}}{W_{2}^{2}}\right) & -A_{1}\left(\frac{\alpha_{11}\beta_{11}\beta_{12}}{W_{1}^{2}}+\frac{\alpha_{21}\beta_{21}\beta_{22}}{W_{2}^{2}}\right)\\
-A_{2}\left(\frac{\alpha_{12}\beta_{11}\beta_{12}}{W_{1}^{2}}+\frac{\alpha_{22}\beta_{21}\beta_{22}}{W_{2}^{2}}\right) & G_{2}-A_{2}\left(\frac{\alpha_{12}\beta_{12}^{2}}{W_{1}^{2}}+\frac{\alpha_{22}\beta_{22}^{2}}{W_{2}^{2}}\right)
\end{array}\right]\label{eq:jacobian_pollinators}
\end{equation}

\noindent where

\begin{align}
G_{1}(A_{1},A_{2}) & =\frac{\alpha_{11}\beta_{11}}{w_{1}+\beta_{11}A_{1}+\beta_{12}A_{2}}+\frac{\alpha_{21}\beta_{21}}{w_{2}+\beta_{21}A_{1}+\beta_{22}A_{2}}-d_{1}\label{eq:gr1}\\
G_{2}(A_{1},A_{2}) & =\frac{\alpha_{12}\beta_{12}}{w_{1}+\beta_{11}A_{1}+\beta_{12}A_{2}}+\frac{\alpha_{22}\beta_{22}}{w_{2}+\beta_{21}A_{1}+\beta_{22}A_{2}}-d_{2}\label{eq:gr2}
\end{align}

\noindent and

\begin{align}
W_{1}(A_{1},A_{2}) & =w_{1}+\beta_{11}A_{1}+\beta_{12}A_{2}\label{eq:bigW1}\\
W_{2}(A_{1},A_{2}) & =w_{2}+\beta_{21}A_{1}+\beta_{22}A_{2}.\label{eq:bigW2}
\end{align}

\noindent At the trivial equilibrium the jacobian is diagonal and its eigenvalues
are $\lambda_{1}=G_{1}(0,0)$ and $\lambda_{2}=G_{2}(0,0)$. Thus, the trivial equilibrium
is unstable if any of (\ref{eq:viable1}) or (\ref{eq:viable2}) hold. At the A1-only
equilibrium $G_{1}(A_{1},0)=0$ and the eigenvalues are

\noindent 
\[
\lambda_{1}=-A_{1}\left(\frac{\alpha_{11}\beta_{11}^{2}}{W_{1}^{2}}+\frac{\alpha_{21}\beta_{21}^{2}}{W_{2}^{2}}\right)\,,\,\lambda_{2}=G_{2}(A_{1},0).
\]

\noindent Thus, stability depends on the sign of $G_{2}(A_{1},0)$. If $G_{2}(A_{1},0)<0$
pollinator 1 is stable against invasion by pollinator 2, if $G_{2}(A_{1},0)>0$ pollinator
1 can be invaded by pollinator 2. $G_{2}(A_{1},0)$ can be evaluated explicitly,
but the resulting expression is quite complex and we do not give it here. By symmetry,
the A2-only equilibrium is stable against invasion by pollinator 1 if $G_{1}(0,A_{2})<0$
and unstable if $G_{1}(0,A_{2})>0$.

Provided that the coexistence equilibrium exists (i.e., $\hat{A}_{1}>0,\hat{A}_{2}>0$
in \ref{eq:pol-equilibrium}), then $G_{1}(\hat{A}_{1},\hat{A}_{2})=G_{2}(\hat{A}_{1},\hat{A}_{2})=0$
by definition. Thus the trace of the jacobian is negative, which means that stability
depends on the sign of the jacobian determinant, which is

\begin{equation}
\Delta=\frac{\hat{A}_{1}\hat{A}_{2}}{(W_{1}W_{2})^{2}}(\beta_{11}\beta_{22}-\beta_{12}\beta_{21})(\alpha_{11}\alpha_{22}\beta_{11}\beta_{22}-\alpha_{12}\alpha_{21}\beta_{12}\beta_{21}).\label{eq:jacdet}
\end{equation}

\noindent If $\Delta>0$ the equilibrium is locally stable, if $\Delta<0$ it is
unstable. If we replace back the definitions of $\alpha$'s and $\beta$'s in (\ref{eq:jacdet})
the stability condition reads

\[
(u_{1}b_{11}v_{2}b_{22}-v_{1}b_{12}u_{2}b_{21})(e_{11}e_{22}u_{1}b_{11}v_{2}b_{22}-e_{12}e_{21}v_{1}b_{12}u_{2}b_{21})>0.
\]

The above results can be used to study coexistence of specialized pollinators. First,
we consider specialized pollinators pollinating a single plant. For example, let
us assume that both pollinators pollinate plant P1 only, i.e., $u_{1}=v_{1}=1$.
Then $\beta_{11}=\beta_{12}=0$ and substituting these values in (\ref{eq:pol-subsystem})
shows that the two isoclines are parallel lines, i.e., generically, there is no equilibrium.
The same conclusion holds in the case where both pollinators specialize on plant
P2. Thus, two specialist pollinators cannot survive on a single plant.

Second, we consider two pollinators that specialize on different plants (either $\beta_{12}=\beta_{21}=0$
or $\beta_{11}=\beta_{22}=0$). For example, when $\beta_{12}=\beta_{21}=0$ the
interior equilibrium (\ref{eq:pol-equilibrium}) is

\[
(\hat{A}_{1},\hat{A}_{2})=\left(\frac{\alpha_{11}}{d_{1}}-\frac{w_{1}}{\beta_{11}},\frac{\alpha_{22}}{d_{2}}-\frac{w_{2}}{\beta_{22}}\right)
\]

\noindent and stability condition (\ref{eq:jacdet}) holds. The case where $\beta_{11}=\beta_{22}=0$
is similar.

\section{ESS and Nash equilibria\label{sec:appESS}}

Throughout this appendix we assume that inequality (\ref{eq:pollinator_ess_condition})
holds. From (\ref{eq:fitnessA1mutant}) and (\ref{eq:fitnessA2mutant}), for a given
pollinator distribution $(u_{1},v_{1})\in[0,1]\times[0,1]$, pollinator A1 payoffs
when pollinating exclusively plant P1 or plant P2 are

\[
V_{1}(u_{1},v_{1})=\frac{a_{1}e_{11}b_{11}P_{1}}{w_{1}+u_{1}b_{11}A_{1}+v_{1}b_{12}A_{2}},\;\;\;V_{2}(u_{1},v_{1})=\frac{a_{2}e_{21}b_{21}P_{2}}{w_{2}+(1-u_{1})b_{21}A_{1}+(1-v_{1})b_{22}A_{2}}.
\]
Similarly, pollinator 2 payoffs are

\[
W_{1}(u_{1},v_{1})=\frac{a_{1}e_{12}b_{12}P_{1}}{w_{1}+u_{1}b_{11}A_{1}+v_{1}b_{12}A_{2}},\;\;\;W_{2}(u_{1},v_{1})=\frac{a_{2}e_{22}b_{22}P_{2}}{w_{2}+(1-u_{1})b_{21}A_{1}+(1-v_{1})b_{22}A_{2}}.
\]

First, we consider ESS at which both pollinators are specialists. We start with the
case where both pollinators specialize on plant 1. Strategy $(u_{1},v_{1})=(1,1)$
is an ESS provided $V_{1}(1,1)>V_{2}(1,1)$ and $W_{1}(1,1)>W_{2}(1,1)$. These inequalities
are equivalent to $P_{2}<\frac{a_{1}b_{11}e_{11}w_{2}}{a_{2}b_{21}e_{21}(w_{1}+b_{11}A_{1}+b_{12}A_{2})}P_{1}$
and $P_{2}<Q_{d}P_{1}$, where $Q_{d}$ is given in (\ref{eq:isoleg_d}). Inequality
(\ref{eq:pollinator_ess_condition}) implies that $Q_{d}<\frac{a_{1}b_{11}e_{11}w_{2}}{a_{2}b_{21}e_{21}(w_{1}+b_{11}A_{1}+b_{12}A_{2})}$.
Consequently, for $P_{2}<Q_{d}P_{1}$ strategy $(u_{1},v_{1})=(1,1)$ is the ESS.
Now we consider the case where pollinator 1 specializes on plant 1 and pollinator
2 on plant 2. Strategy $(u_{1},v_{1})=(1,0)$ is an ESS provided $V_{1}(1,0)>V_{2}(1,0)$
and $W_{2}(1,0)>W_{1}(1,0)$. These inequalities are equivalent to $Q_{b}P_{1}>P_{2}>Q_{c}P_{1}$
where $Q_{b}$ is given in (\ref{eq:isoleg_b}). Now we consider the case where both
pollinators specialize on plant 2. Strategy $(u_{1},v_{1})=(0,0)$ is an ESS provided
$V_{2}(0,0)>V_{1}(0,0)$ and $W_{2}(0,0)>W_{1}(0,0)$. These inequalities are equivalent
to $P_{2}>Q_{a}P_{1}$ where $Q_{a}$ is given in (\ref{eq:isoleg_d}) and $P_{2}>\frac{a_{1}b_{12}e_{12}(w_{2}+b_{21}A_{1}+b_{22}A_{2})}{a_{2}b_{22}e_{22}w_{1}}P_{1}$.
Inequality (\ref{eq:pollinator_ess_condition}) implies that $Q_{a}>\frac{a_{1}b_{12}e_{12}(w_{2}+b_{21}A_{1}+b_{22}A_{2})}{a_{2}b_{22}e_{22}P_{2}w_{1}}$.
Consequently, for $P_{2}>Q_{a}P_{1}$ strategy $(u_{1},v_{1})=(0,0)$ is the ESS.
Now we consider the case where pollinator 1 specializes on plant 2 and pollinator
2 on plant 1. Strategy $(u_{1},v_{1})=(0,1)$ is an ESS provided $V_{2}(0,1)>V_{1}(0,1)$
and $W_{1}(0,1)>W_{2}(0,1)$. These inequalities are equivalent to $P_{2}>Q_{b}P_{1}$
and $P_{2}<Q_{c}P_{1}.$ Inequality (\ref{eq:pollinator_ess_condition}) implies
that $Q_{b}>Q_{c}$. Consequently, $(u_{1},v_{1})=(0,1)$ is never an ESS.

Second, we consider ESSs when the first pollinator is a generalist while the second
pollinator is a specialist. Let us assume that the second pollinator specializes
on plant 2, i.e., we seek ESS in the form $(u_{1},0)$ where $0<u_{1}<1$. Such a
strategy must satisfy $V_{1}(u_{1},0)=V_{2}(u_{1},0)$ and $W_{2}(u_{1},0)>W_{1}(u_{1},0)$.
Equality $V_{1}(u_{1},0)=V_{2}(u_{1},0)$ leads to

\[
u_{1}^{*}=\frac{a_{1}b_{11}e_{11}P_{1}(w_{2}+b_{21}A_{1}+b_{22}A_{2})-a_{2}b_{21}e_{21}P_{2}w_{1}}{A_{1}b_{11}b_{21}(a_{1}e_{11}P_{1}+a_{2}e_{21}P_{2})}.
\]

\noindent This value is between 0 and 1 provided $Q_{b}P_{1}<P_{2}<Q_{q}P_{1}$.
Inequality (\ref{eq:pollinator_ess_condition}) implies that $W_{2}(u_{1}^{*},0)>W_{1}(u_{1}^{*},0)$.
Thus, $(u_{1}^{*},0)$ is a Nash equilibrium. To prove it is also an ESS, we need
to verify its stability. Because functions $V_{i}$ ($i=1,2$) are non-linear in
$u_{1}$, we use the local ESS condition \citep{hofbauersigmund1998} $u_{1}^{*}V_{1}(u_{1},0)+u_{2}^{*}V_{2}(u_{1},0)>u_{1}V_{1}(u_{1},0)+u_{2}V_{2}(u_{1},0)$
for every $(u_{1},u_{2})$ ($u_{1}+u_{2}=1$, $u_{1}>0$, $u_{2}>1$) close to (but
different from) $(u_{1}^{*},u_{2}^{*}).$ This condition is equivalent to

\begin{equation}
\frac{(a_{2}b_{11}e_{11}P_{1}(A_{2}b_{22}+A_{1}(b_{21}-b_{21}u_{1})+w_{2}))^{2}}{A_{1}b_{11}b_{21}(a_{1}e_{11}P_{1}+a_{2}e_{21}P_{2})(A_{1}b_{11}u_{1}+w_{1})(A_{2}b_{22}+A_{1}(b_{21}-b_{21}u_{1})+w_{2})}>0.\label{esscondition1}
\end{equation}

\noindent The numerator is positive and the denominator equals to 0 for $u_{1}=-\frac{w_{1}}{A_{1}b_{11}}<0$
and $u_{1}=\frac{A_{1}b_{21}+A_{2}b_{22}+w_{2}}{A_{1}b_{21}}>1.$ Because the denominator
is a quadratic function and its graph is an upside down parabola, inequality (\ref{esscondition1})
holds for all $0<u_{1}<1$. This shows that $(u_{1},v_{1})=(u_{1}^{*},0)$ is an
ESS.

Now we assume that the second pollinator specializes on plant 1, i.e., we seek ESS
in the form $(u_{1},1)$ where $0<u_{1}<1$. Such a strategy must satisfy $V_{1}(u_{1},1)=V_{2}(u_{1},1)$
and $W_{1}(u_{1},1)>W_{2}(u_{1},1)$. The equality leads to

\[
u_{1}^{*}=\frac{a_{1}b_{11}e_{11}P_{1}(w_{2}+b_{21}A_{1})-a_{2}b_{21}e_{21}P_{2}(w_{1}+b_{12}A_{2})}{A_{1}b_{11}b_{21}(a_{1}e_{11}P_{1}+a_{2}e_{21}P_{2})}.
\]

\noindent Then

\begin{align*}
W_{1}(u_{1}^{*},1) & =\frac{b_{12}b_{21}e_{12}(a_{1}e_{11}P_{1}+a_{2}e_{21}P_{2})}{b_{21}e_{11}(A_{1}b_{11}+A_{2}b_{12}+w_{1})+b_{11}e_{11}w_{2}}\\
W_{2}(u_{1}^{*},1) & =\frac{b_{11}b_{22}e_{22}(a_{1}e_{11}P+a_{2}e_{21}P_{2})}{b_{21}e_{21}(A_{1}b_{11}+A_{2}b_{12}+w_{1})+b_{11}e_{21}w_{2}}
\end{align*}

\noindent and inequality (\ref{eq:pollinator_ess_condition}) implies that $W_{2}(u_{1}^{*},1)>W_{1}(u_{1}^{*},1)$
and thus $(u_{1}^{*},1)$ is never an ESS.

Third, we consider ESSs when the first pollinator is a specialist while the second
pollinator is a generalist. Let us assume that the first pollinator specializes on
plant P1, i.e., we seek ESS in the form $(1,v_{1})$ where $0<v_{1}<1$. Such a strategy
must satisfy $V_{1}(1,v_{1})>V_{2}(1,v_{1})$ and $W_{1}(1,v_{1})=W_{2}(1,v_{1})$.
The equality leads to

\[
v_{1}^{*}=\frac{a_{1}b_{12}e_{12}P_{1}(w_{2}+b_{22}A_{2})-a_{2}b_{22}e_{22}P_{2}(w_{1}+b_{11}A_{1})}{A_{2}b_{12}b_{22}(a_{1}e_{12}P_{1}+a_{2}e_{22}P_{2})}.
\]

\noindent This value is between 0 and 1 provided $Q_{d}P_{1}<P_{2}<Q_{c}P_{1}$.
Inequality (\ref{eq:pollinator_ess_condition}) implies that $V_{1}(1,v_{1}^{*})>V_{2}(1,v_{1}^{*})$.
The local ESS condition requires $v_{1}^{*}W_{1}(1,v_{1})+v_{2}^{*}W_{2}(1,v_{1})>v_{1}W_{1}(1,v_{1})+v_{2}W_{2}(1,v_{1})$
for every $(v_{1},v_{2})$ ($v_{1}+v_{2}=1$, $v_{1}>0$, $v_{2}>1$) close to (but
different from) $(v_{1}^{*},v_{2}^{*}).$ This condition is equivalent to

\begin{equation}
\frac{(a_{1}b_{12}e_{12}P_{1}(A_{2}b_{22}(v_{1}-1)-w_{2})+a_{2}b_{22}e_{22}P_{2}(A_{1}b_{11}+A_{2}b_{12}v_{1}+w_{1}))^{2}}{A_{2}b_{12}b_{22}(w_{2}-A_{2}b_{22}(v_{1}-1))(a_{1}e_{12}P_{1}+a_{2}e_{22}P_{2})(A_{1}b_{11}+A_{2}b_{12}v_{1}+w_{1})}>0.\label{esscondition2}
\end{equation}

\noindent The numerator is positive and the denominator equals to 0 for $v_{1}=-\frac{A_{1}b_{11}+w_{1}}{A_{2}b_{12}}<0$
and $v_{1}=\frac{w_{2}}{A_{2}b_{22}}+1>1.$ Because the denominator is a quadratic
function and its graph is an upside down parabola, inequality (\ref{esscondition2})
holds for all $0<v_{1}<1$. This shows that $(u_{1},v_{1})=(1,v_{1}^{*})$ is an
ESS.

Now we assume that the first pollinator specializes on plant P2, i.e., we seek ESS
in the form $(0,v_{1})$ where $0<v_{1}<1$. Such a strategy must satisfy $V_{2}(0,v_{1})>V_{1}(0,v_{1})$
and $W_{1}(0,v_{1})=W_{2}(0,v_{1})$. The equality leads to

\[
v_{1}^{*}=\frac{a_{1}b_{12}e_{12}P_{1}(w_{2}+b_{21}A_{1}+b_{22}A_{2})-a_{2}b_{22}e_{22}P_{2}w_{1}}{A_{2}b_{12}b_{22}(a_{1}e_{12}P_{1}+a_{2}e_{22}P_{2})}.
\]

\noindent However, inequality (\ref{eq:pollinator_ess_condition}) implies that $V_{1}(0,v_{1}^{*})>V_{2}(0,v_{1}^{*})$
so that no ESS in the form $(0,v_{1})$ exists.

Fourth, we consider the case where both pollinators are generalists. This situation
corresponds to ESS of the form $(u_{1},v_{1})$ with $0<u_{1}<1$ and $0<v_{1}<1$.
Such an ESS must satisfy $V_{1}(u_{1},v_{1})=V_{2}(u_{1},v_{1})$ and $W_{1}(u_{1},v_{1})=W_{2}(u_{1},v_{1})$.
These equalities are equivalent to

\[
\begin{array}{lll}
a_{1}e_{11}b_{11}P_{1}(w_{2}+u_{2}b_{21}A_{1}+v_{2}b_{22}A_{2}) & = & a_{2}e_{21}b_{21}P_{2}(w_{1}+u_{1}b_{11}A_{1}+v_{1}b_{12}A_{2})\\[0.2cm]
a_{1}e_{12}b_{12}P_{1}(w_{2}+u_{2}b_{21}A_{1}+v_{2}b_{22}A_{2}) & = & a_{2}e_{22}b_{22}P_{2}(w_{1}+u_{1}b_{11}A1+v_{1}b_{12}A_{2}).
\end{array}
\]

\noindent Because $w_{1}+u_{1}b_{11}A_{1}+v_{1}b_{12}A_{2}>0$ and $w_{2}+u_{2}b_{21}A_{1}+v_{2}b_{22}A_{2}>0$,
inequality (\ref{eq:pollinator_ess_condition}) implies that these two equations
do not have any solution $(u_{1},v_{1})\in[0,1]\times[0,1]$. Thus, it is impossible
for both pollinators to be generalists.

\section{Plant dynamics in regions I, III and V\label{sec:appP1P2regions}}

First we calculate plant P1 boundary equilibrium. From Table \ref{tab:regions_plaspace}
it follows that this equilibrium is in region V where both pollinators pollinate
P1. Substituting $(u_{1},v_{1})=(1,1)$ in (\ref{eq:ode_p1}) and (\ref{eq:ode_p2}),
and solving for equilibria when $P_{1}\neq0$ and $P_{2}=0$ leads to equilibrium
(\ref{eq:P1boundary}).

The plant population dynamics in region V are

\begin{align*}
\frac{dP_{1}}{dt} & =\left(\frac{a_{1}(r_{11}b_{11}A_{1}+r_{12}b_{12}A_{2})}{w_{1}+b_{11}A_{1}+b_{12}A_{2}}\left(1-\frac{P_{1}+c_{2}P_{2}}{K_{1}}\right)-m_{1}\right)P_{1}\\
\frac{dP_{2}}{dt} & =-m_{2}P_{2},
\end{align*}

\noindent and provided plant 1 is viable (i.e., (\ref{viab1}) holds), equilibrium
(\ref{eq:P1boundary}) exists (is positive) and is locally asymptotically stable.
Following the same steps above \emph{mutatis mutandis}, leads to equation (\ref{eq:P2boundary})
for plant P2 boundary equilibrium in region I (where ESS is $(u_{1},v_{1})=(0,0)$,
see Table \ref{tab:regions_plaspace}). Analogously, if (\ref{viab2}) holds then
(\ref{eq:P2boundary}) exists and is locally asymptotically stable.

Now we consider plant population dynamics in region III. According to Table \ref{tab:regions_plaspace}
the ESS strategy in this region is $(u_{1},v_{1})=(1,0)$, i.e., pollinator A1 (A2)
interacts only with plant P1 (P2). Substituting these preferences in (\ref{eq:ode_p1})
and (\ref{eq:ode_p2}), plant population dynamics in region III are described by
the Lotka\textendash Volterra competition model

\begin{align*}
\frac{dP_{1}}{dt} & =(s_{1}-m_{1})P_{1}\left(1-\frac{P_{1}+c_{2}P_{2}}{H_{1}}\right)\\
\frac{dP_{2}}{dt} & =(s_{2}-m_{2})P_{2}\left(1-\frac{P_{2}+c_{1}P_{1}}{H_{2}}\right),
\end{align*}
where

\begin{align*}
s_{1} & =\frac{a_{1}r_{11}b_{11}A_{1}}{w_{1}+b_{11}A_{1}},\;\;s_{2}=\frac{a_{2}r_{22}b_{22}A_{2}}{w_{2}+b_{22}A_{2}},\;\;H_{1}=K_{1}\left(1-\frac{m_{1}}{s_{1}}\right),\;\;H_{2}=K_{2}\left(1-\frac{m_{2}}{s_{2}}\right).
\end{align*}
Plant population dynamics in region III depend on the position of plant isoclines

\begin{align*}
P_{1}+c_{2}P_{2} & =H_{1}\\
P_{2}+c_{1}P_{1} & =H_{2}.
\end{align*}
Provided the plant isoclines intersect in region III, the coexistence equilibrium
is

\begin{equation}
(\hat{P}_{1},\hat{P}_{2})=\left(\frac{H_{1}-c_{2}H_{2}}{1-c_{1}c_{2}},\frac{H_{2}-c_{1}H_{1}}{1-c_{1}c_{2}}\right).\label{eq:P1P2intIII}
\end{equation}
For (\ref{eq:P1P2intIII}) to be in region III, it must satisfy $Q_{b}\hat{P}_{1}<\hat{P}_{2}<Q_{c}\hat{P}_{1}$,
where $Q_{b}$ and $Q_{c}$ are given in (\ref{eq:isoleg_b}) and (\ref{eq:isoleg_c}),
respectively. Substituting $Q_{b}$ , $Q_{c}$ and (\ref{eq:P1P2intIII}) we get

\begin{equation}
\frac{b_{21}e_{21}}{b_{11}e_{11}}<\frac{a_{1}(w_{2}+b_{22}A_{2})(H_{1}-c_{2}H_{2})}{a_{2}(w_{1}+b_{11}A_{1})(H_{2}-c_{1}H_{1})}<\frac{b_{22}e_{22}}{b_{12}e_{12}}.\label{eq:P1P2intIIIfeasible}
\end{equation}

Provided (\ref{eq:P1P2intIIIfeasible}) holds, the local stability of (\ref{eq:P1P2intIII})
depends on the competition coefficients. From the Lotka\textendash Volterra theory
(\ref{eq:P1P2intIII}) is locally stable if $c_{1}c_{2}<1$. If $c_{1}c_{2}>1$,
(\ref{eq:P1P2intIII}) is unstable, and trajectories will approach either isoleg-b
and cross into region II, or approach isoleg-c and cross into region IV depending
on the initial conditions. If (\ref{eq:P1P2intIIIfeasible}) does not hold there
is no plant equilibrium in region III.

In the special case where $c_{1}=c_{2}=0$, the stable equilibrium in region III
is

\[
(\hat{P}_{1},\hat{P}_{2})=(H_{1},H_{2})=\bigg(\frac{K_{1}(A_{1}b_{11}(a_{1}r_{11}-m_{1})-m_{1}w_{1})}{a_{1}A_{1}b_{11}r_{11}},\frac{K_{2}(A_{2}b_{22}(a_{2}r_{22}-m_{2})-m_{2}w_{2})}{a_{2}A_{2}b_{22}r_{22}}\bigg).
\]
At this equilibrium plant 1 density is higher than is the plant density at the boundary
equilibrium (\ref{eq:P1boundary}) in region V iff

\[
\frac{r_{11}}{r_{12}}>1+\frac{w_{1}}{b_{11}A_{1}}.
\]
This shows that provided $r_{11}>r_{12}$ and pollinator A1 is abundant enough, plant
P1 density at the interior equilibrium in region III will be higher than is the plant
P1 density at the boundary equilibrium. Analogous conclusions apply to plant P2.

\section{Coexistence conditions for pollinators with adaptive preferences\label{sec:appA1A2coexflex}}

Using Table \ref{tab:regions_plaspace} in the main text, we rewrite isolegs characterizing
regions I-V in terms of pollinator densities. For example, isoleg-a that separates
regions I and II in the plant plane $P_{1}P_{2}$ is given by equation $\frac{P_{2}}{P_{1}}=Q_{a}(A_{1},A_{2})=\frac{a_{1}b_{11}e_{11}(w_{2}+b_{21}A_{1}+b_{22}A_{2})}{a_{2}b_{21}e_{21}w_{1}}.$
Solving for $A_{2}$ leads to isoleg-a in the pollinator plane

\[
A_{2}=S_{a}A_{1}+I_{a}=-\frac{b_{21}}{b_{22}}A_{1}+\frac{a_{2}b_{21}e_{21}P_{2}w_{1}-a_{1}b_{11}e_{11}P_{1}w_{2}}{a_{1}b_{11}b_{22}e_{11}P_{1}},
\]
and the other isolegs in the pollinator plane $A_{1}A_{2}$ are obtained analogously
and they are listed in Table \ref{tab:regions_polspace} in the main text.

Isoleg-b and isoleg-c (which enclose region III) do not intersect in the positive
part of the $A_{1}A_{2}$ plane. Indeed, the intersection point is $A_{1}=(I_{c}-I_{b})/(S_{b}-S_{c}).$
From (\ref{eq:slopes_intercepts}), $I_{c}-I_{b}=\frac{a_{2}P_{2}w_{1}}{a_{1}P_{1}b_{2}}\left(\frac{e_{22}b_{22}}{e_{12}b_{12}}-\frac{e_{21}b_{21}}{e_{11}b_{11}}\right)$
and $S_{b}-S_{a}=\frac{a_{2}P_{2}}{a_{1}P_{1}}\left(\frac{e_{21}b_{21}}{e_{11}b_{11}}-\frac{e_{22}b_{22}}{e_{12}b_{12}}\right)$
have different signs, thus isoleg-b and isoleg-c intersection is non-positive.

Now we show that for given parameters and plant population densities it is not possible
that both regions I and V co-exist. We observe that region I exists in the positive
quadrant iff $I_{a}>0$, because in this case isoleg-a intersects both A1 and A2
axes at positive values. Similarly, region V exists in the positive quadrant iff
$I_{d}>0$, because in this case isoleg-d intersects both A1 and A2 axes at positive
values. However, condition (\ref{eq:pollinator_ess_condition}) rules out the possibility
that both $I_{a}$ and $I_{d}$ are positive.

We determine conditions for pollinator coexistence in regions I to V. Appendix \ref{sec:appA1A2coex}
shows that two pollinators specialized on the same plant (both $u_{1}$ and $v_{1}$
equal to 0 or 1) cannot coexist. This rules out coexistence in regions I and V. Now,
let us consider region II where pollinator A1 is a generalist and A2 specializes
on plant P2. Thus $0<u_{1}=u_{1}^{*}<1$, $v_{1}=0$, and the A2 isocline is

\begin{equation}
\frac{a_{2}e_{22}b_{22}P_{2}}{w_{2}+(1-u_{1}^{*})b_{21}A_{1}+b_{22}A_{2}}=d_{2}.\label{eq:nullA1reg2}
\end{equation}

\noindent Because the payoff of pollinator A1 when pollinating plant P1 is the same
as when pollinating plant P2, the A1 isocline is

\noindent 
\begin{equation}
\frac{a_{2}e_{21}b_{21}P_{2}}{w_{2}+(1-u_{1}^{*})b_{21}A_{1}+b_{22}A_{2}}=d_{1}.\label{eq:nullA2reg2}
\end{equation}

\noindent Substituting $u_{1}^{*}$ (\ref{eq:u_star}) in (\ref{eq:nullA1reg2})
and (\ref{eq:nullA2reg2}) shows that both these equalities define parallel lines
in pollinator phase space. Thus, generically, there cannot be a coexistence equilibrium
in region II. Pollinator A2 will displace A1 if

\noindent 
\[
\frac{e_{22}b_{22}}{e_{21}b_{21}}>\frac{d_{2}}{d_{1}},
\]

\noindent or A1 will displace A2 if the opposite inequality holds.

In region IV, pollinator A2 is a generalist and A1 specializes on plant P1. Because
of symmetry, the last result applies \emph{mutatis mutandis}. This means that either
A1 will displace A2 if

\[
\frac{e_{11}b_{11}}{e_{12}b_{12}}>\frac{d_{1}}{d_{2}},
\]
or A2 will displace A1 if the opposite inequality holds. If

\begin{equation}
\frac{e_{11}b_{11}}{e_{12}b_{12}}>\frac{d_{1}}{d_{2}}>\frac{e_{21}b_{21}}{e_{22}b_{22}},\label{eq:coexA1A2stable}
\end{equation}
then A1 can be invaded by A2 and vice versa and coexistence by mutual invasion occurs.

Finally, let us consider region III, in which pollinator A1 specializes in plant
P1 $(u_{1}=1)$, and A2 specializes in P2 $(v_{1}=0)$. The pollinator isoclines
intersect at

\begin{equation}
(\hat{A}_{1},\hat{A}_{2})=\left(\frac{a_{1}e_{11}P_{1}}{d_{1}}-\frac{w_{1}}{b_{11}},\frac{a_{2}e_{22}P_{2}}{d_{2}}-\frac{w_{2}}{b_{22}}\right).\label{eq:A1A2intIII}
\end{equation}
To be a coexistence equilibrium however, $(\hat{A}_{1},\hat{A}_{2})$ must lie between
isolegs $S_{b}A_{1}+I_{b}$ and $S_{c}A_{1}+I_{c}$, i.e., $S_{b}\hat{A}_{1}+I_{b}<\hat{A}_{2}<S_{c}\hat{A}_{1}+I_{c}.$
Substituting (\ref{eq:A1A2intIII}) in these inequalities leads to

\[
\frac{b_{21}e_{21}}{b_{22}e_{22}}<\frac{d_{1}}{d_{2}}<\frac{b_{11}e_{11}}{b_{12}e_{12}},
\]
which are the conditions for mutual invasion (\ref{eq:coexA1A2stable}). Thus, a
coexistence equilibrium within region III is locally stable. Since there are no coexistence
equilibria within regions II and IV, we conclude that if (\ref{eq:coexA1A2stable})
holds there is a single stable coexistence equilibrium within region III.

\section{Combined plant\textendash pollinator dynamics\label{sec:dynamics_module}}

\begin{figure}[t]
\begin{centering}
\includegraphics[width=1\textwidth]{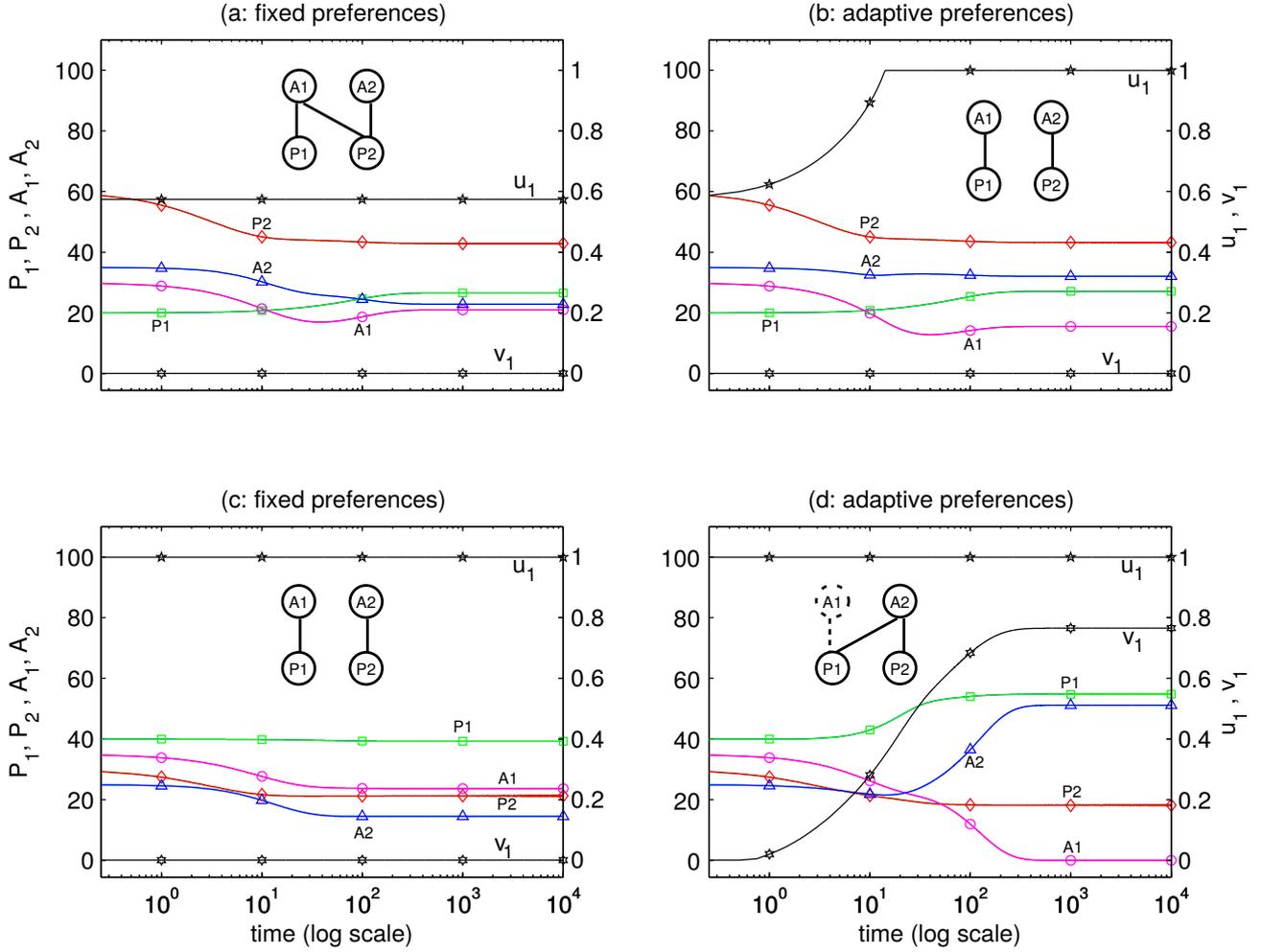} 
\par\end{centering}
\caption{\label{fig:dynamics}Dynamics of system (\ref{eq:ode_system}). Preferences given
according to: (a,c) ESS at $t=0$ and kept fixed for $t>0$; (b,d) ESS at all times
$(t\geq0)$. Densities (left axes) represented by $P_{1}$: green squares; $P_{2}$:
red diamonds; $A_{1}$: pink circles; $A_{2}$: blue triangles. Plant 1 preferences
(right axes) represented by $u_{1}$: 5-pointed, $v_{1}$: 6-pointed stars. Inset
graphs display final plant\textendash pollinator interactions (dash stroke for extinct
species). Parameters (a,b,c,d): $r_{i}=0.1$, $m_{i}=0.01$, $b_{ij}=0.1$, $a_{i}=0.4$,
$w_{i}=0.25$, $e_{11}=e_{22}=0.2$, $d_{1}=0.12,d_{2}=0.1$; (a,b): $K_{i}=50$,
$e_{21}=0.17,e_{12}=0.1$; (c,d): $K_{1}=60,K_{2}=30$, $e_{12}=0.17,e_{21}=0.1$.
Competition coefficients (a,b): $c_{i}=0.2$; (c,d): $c_{1}=0.4,c_{2}=0.1$.}
\end{figure}

Figure \ref{fig:dynamics} illustrates the population dynamics of the four species
system (\ref{eq:ode_system}) when pollinator preferences for plants are fixed (left
panels) or adaptive (right panels). Panels in each row assume the same parameters
and initial conditions. Initial preferences $u_{1}(0),v_{1}(0)$ are calculated as
the ESS (Table \ref{tab:regions_plaspace} or \ref{tab:regions_polspace}) for the
initial densities $P_{1}(0),P_{2}(0),A_{1}(0),A_{2}(0).$ In the left column of Figure
\ref{fig:dynamics} these preferences are kept fixed at their initial values (their
time series remain horizontal), while in the right column preferences track changes
in population densities (ESS) instantaneously.

In the top row (panels a,b) pollinator A1 starts as a generalist biased towards plant
P1 $(u_{1}\approx0.6)$, and A2 as a P2 specialist $(v_{1}=0)$. In panel (a) these
preferences remain fixed and all four species attain stable coexistence. In panel
(b) preferences adapt and the four species attain coexistence again, but pollinator
A1 turns into a plant P1 specialist. Here adaptation leads to the end of competition
between A1 and A2, which do not share any plant.

The bottom row (panels c, d) uses a different parameter set, and the initial conditions
make pollinator A1 a plant P1 specialist $(u_{1}=1)$ and A2 a P2 specialist $(v_{1}=0)$.
Thus, A1 and A2 do not compete initially, and four species coexistence happens if
preferences remain fixed (c). If preferences adapt, panel (d) shows that pollinator
A2 becomes a generalist. As the preference for P1 grows larger for A2, strong competition
drives specialist pollinator A1 towards extinction.

\end{document}